\documentclass[useAMS,usegraphicx]{mn2e}
\usepackage{lscape}
\usepackage{latexsym}
\usepackage{epsfig}
\begin{document}

\title[A complete sample of Swift short GRBs]{A complete sample of bright {\it Swift} short Gamma--Ray Bursts}

\author[P. D'Avanzo et al.]{P. D'Avanzo$^{1}$\thanks{E-mail: paolo.davanzo@brera.inaf.it}, R. Salvaterra$^{2}$, M. G. Bernardini$^{1}$, L. Nava$^{3}$, S. Campana$^{1}$, S. Covino$^{1}$, \newauthor V. D'Elia$^{4}$, G. Ghirlanda$^{1}$, G. Ghisellini$^{1}$, A. Melandri$^{1}$, B. Sbarufatti$^{1,5}$, S. D. Vergani$^{6,1}$ \newauthor \& G. Tagliaferri$^{1}$ \\
$^{1}$INAF - Osservatorio Astronomico di Brera, via E. Bianchi 46, I-23807, Merate (LC), Italy\\
$^{2}$INAF - IASF Milano, via E. Bassini 15, I-20133, Milano, Italy\\
$^{3}$The Racah Institute of physics, The Hebrew University of Jerusalem, Jerusalem 91904, Israel\\
$^{4}$ASI - Science Data Centre, via G. Galilei, I-00044, Frascati, Italy\\
$^{5}$Department of Astronomy and Astrophysics, Pennsylvania State University, University Park, PA, 16802, USA\\
$^{6}$GEPI, Observatoire de Paris, CNRS, Univ. Paris Diderot, 5 place Jules Janssen, F-92190 Meudon, France\\}

\def\simg{\mathrel{%
      \rlap{\raise 0.511ex \hbox{$>$}}{\lower 0.511ex \hbox{$\sim$}}}}
\def\siml{\mathrel{%
      \rlap{\raise 0.511ex \hbox{$<$}}{\lower 0.511ex \hbox{$\sim$}}}}
\def\Mesz{M\'esz\'aros~}
\def\ie{i.e$.$~} \def\eg{e.g$.$~} \def\etal{et al$.$~} 
\def\eq{eq$.$~} \def\eqs{eqs$.$~} \def\deg{^{\rm o}} \def\dd{{\rm d}}
\def\beq{\begin{equation}} \def\eeq{\end{equation}}
\def\epsel{\varepsilon_e} \def\epse1{\varepsilon_{e,-1}} 
\def\epsmag{\varepsilon_B} \def\eBzero{\tilde{\varepsilon_B}} 
\def\eBone{\varepsilon_{B,-1}} \def\eBtwo{\varepsilon_{B,-2}} 
\def\eBthree{\varepsilon_{B,-3}} \def\eBfour{\varepsilon_{B,-4}}
\def\E53{E_{53}} \def\nex0{n_{*,0}} \def\Gm02{\Gamma_{0,2}} 
\def\D28{D_{28}^{-2}} \def\nuo{\nu_{14.6}}
\def\Tdone{T_{d,-1}} \def\Tdtwo{T_{d,-2}} \def\Tdthree{T_{d,-3}}

\date{Accepted . Received ; }

\pagerange{\pageref{firstpage}--\pageref{lastpage}} \pubyear{2014}

\label{firstpage}

\maketitle

\begin{abstract}

We present a carefully selected sample of short gamma-ray bursts (SGRBs) observed by the {\it Swift} satellite up to June 2013. Inspired by the criteria we used to build a similar sample of bright long GRBs (the BAT6 sample), we selected SGRBs with favorable observing conditions for the redshift determination on ground, ending up with a sample of 36 events, almost half of which with a redshift measure. The redshift completeness increases up to about 70\% (with an average redshift value of $z = 0.85$) by restricting to those events that are bright in the 15-150 keV {\it Swift} Burst Alert Telescope energy band. Such flux-limited sample minimizes any redshift-related selection effects, and can provide a robust base for the study of the energetics, redshift distribution and environment of the {\it Swift} bright population of SGRBs. For all the events of the sample we derived the prompt and afterglow emission in both the observer and (when possible) rest frame and tested the consistency with the correlations valid for long GRBs. The redshift and intrinsic X-ray absorbing column density distributions we obtain are consistent with the scenario of SGRBs originated by the coalescence of compact objects in primordial binaries, with a possible minor contribution ($\sim$10\%--25\%) of binaries formed by dynamical capture (or experiencing large natal kicks). This sample is expected to significantly increase with further years of {\it Swift} activity. 

\end{abstract}


\section{Introduction}

Gamma-ray bursts (GRBs) are rapid, powerful flashes of gamma-ray radiation occuring
at an average rate of one event per day over the whole sky at cosmological distances. The high energy prompt emission is followed by a broadband (X-rays to radio ranges) fading emission, named afterglow, (Costa et al. 1997; van Paradijs et al. 1997; Frail et al. 1997; Bremer et al. 1998; Heng et al. 2008) that can be observed up to weeks and months after the onset of the event. 

The distribution of GRB durations observed by the BATSE\footnote{Burst and Transient Source Experiment, on board the {\it Compton Gamma Ray Observatory}} instrument (Fishman et al. 1989) is bimodal, with peaks at $T_{90} \sim0.2$ and $T_{90} \sim20$\,s and a boundary at $T_{90} \sim 2$~s (Kouveliotou et al. 1993)\footnote{$T_{90}$ is defined as the time during which the cumulative counts increase from 5\% to 95\% above background, adding up to 90\% of the total GRB counts.}. The two classes of long (LGRBs, $T_{90} > 2$ s) and short GRBs (SGRBs, $T_{90} \le 2$ s), show substantial evidences for different origins. 
While it has been firmly established that LGRBs, or at least a significant fraction of the nearby events (with redshift $z \leq 1$) for which it has been possible to search for the presence of a supernova (SN), are associated with core-collapse explosions of massive stars (see Hjorth \& Bloom 2011, for a recent review), the nature of SGRB progenitors is still under debate. 
Current models suggest that they are associated with the merging of compact objects in binary systems, such as a double neutron star (NS), or a NS and a black hole (BH) system (Eichler et al. 1989; Narayan et al. 1992; Nakar 2007; Berger 2014). Such systems can originate from the evolution of massive stars in a primordial binary (Narayan et al. 1992) or by dynamical interactions in globular clusters during their core collapse (Grindlay et al. 2006; Salvaterra et al. 2008). Very recently,  a direct evidence supporting the merger scenario has been claimed by Tanvir et al. (2013) and Berger, Fong \& Chornock (2013) who reported the possible detection of a kilonova associated to the SGRB\,130603B (but see Jin et al. 2013 for further discussion).

SGRBs are not distinguished from LGRBs only by their duration. If we consider the observed prompt emission, negligible spectral lag (Norris, Marani \& Bonnell 2000; Norris, Scargle \& Bonnell 2001) and harder spectra (Kouveliotou et al. 1993) are common for SGRBs. On the other hand, the prompt emission properties of short GRBs are similar to the first 1--2 s of long events (Ghirlanda, Ghisellini \& Celotti 2004) and both classes of objects show a similar spectral evolution (Ghirlanda, Ghisellini \& Nava 2011). This might suggest a common emission mechanism for both long and short GRBs. 

Since 2005, with the advent of the fast-repointing {\it Swift} satellite (Gehrels et al. 2004), the discovery of SGRBs afterglows and the identification of their host galaxies made possible to study their distances, energy scales and environments (Gehrels et al. 2005). 
SGRBs are found to be typically less energetic (their isotropic equivalent energy, $E_{\mathrm{iso}}$, is of the order of $10^{49} - 10^{51}\,\mathrm{erg}$) than LGRBs and to occur at a lower redshift (Nakar 2007; Berger 2011; Fong et al. 2013). Their afterglows tend to be significantly fainter on average than those of LGRBs (Kann et al. 2011; Nicuesa Guelbenzu et al. 2012; Margutti et al. 2013), but scaled to their $E_{\mathrm{iso}}$ similar to faint LGRBs (Nysewander et al. 2009). Concerning the host galaxies, SGRBs occur in both early and late type galaxies with low star formation rate and are associated with an old stellar population (Berger 2009; Leibler \& Berger 2010; Fong et al. 2013). 
A different origin for SGRBs with respect to the LGRB class is also supported by the lack of detection of the underlying SN in the light curves of their optical afterglows down to very stringent magnitude limits (Hjorth et al. 2005a; Hjorth et al. 2005b; Fox et al. 2005; Covino et al. 2006; Kann et al. 2011; D'Avanzo et al. 2009) and by their inconsistency with the correlation, valid for LGRBs, between the rest frame spectral peak energy and $E_\mathrm{iso}$ ($E_\mathrm{peak}-E_\mathrm{iso}$ correlation; Amati et al. 2008 and references therein). On the other hand, Ghirlanda et al. (2009) showed that SGRBs are consistent with the same $E_\mathrm{peak}-L_\mathrm{iso}$ correlation (where $L_\mathrm{iso}$ is the prompt emission isotropic peak luminosity) defined by LGRBs (Yonetoku et al. 2004). SGRBs are also consistent with the three parameter $E_\mathrm{iso}-E_\mathrm{peak}-E_\mathrm{X}$ correlation (with $E_\mathrm{X}$ being the afterglow energy emitted in the soft X-ray band; Bernardini et al. 2012; Margutti et al. 2013). Furthermore, the distributions of the intrinsic X-ray absorbing column densities of long and short GRBs do not show significant differences when compared in the same redshift range ($z \leq 1$; Kopac et al. 2012; Margutti et al. 2013). Finally, given that the measured duration of the GRB prompt emission can vary for different instrument (e.g. due to the different energy band used), it has been recently proposed that the value of $T_{90}$ used to divide the long and short GRBs should be reduced to about 0.8 s for the {\it Swift} bursts (Bromberg et al. 2013). A recent review of the properties of SGRBs has been presented by Berger (2014).

The majority of the studies reported above is based over the entire sample of SGRBs with measured redshifts. Although this approach has the clear advantage of describing the intrinsic physical properties of these objects, it can be severely affected by observational biases, given that almost 3/4 of the {\it Swift} SGRBs are lacking a secure redshift measurement (Berger 2014). 

In this paper, we present a carefully selected sub-sample of the {\it Swift} SGRBs. Inspired by the criteria we followed to build a complete sample of LGRBs (the BAT6 sample; Salvaterra et al. 2012 and references therein), we selected SGRBs with favorable observing conditions for redshift determination from the ground, ending up with a sample of 36 events, almost half of which with a redshift measure. The redshift completeness increases up to about 70\% by restricting to those events that are bright in the 15-150 keV {\it Swift} Burst Alert Telescope (BAT; Barthelmy et al. 2005) energy band, obtaining a sample of 16 SGRBs, which is complete in flux and has the highest completeness in redshift with respect to the SGRB samples presented in the literature to date. Such flux-limited sample provides a robust base for the study of the energetics, redshift distribution and environment of the {\it Swift} population of SGRBs. A statistical study of these properties is a useful tool to indirectly investigate their elusive progenitors, find additional parameters for the GRB classification (that can go beyond the duration of the prompt emission) and check for the existence of possible sub-classes.

The paper is organized as follows. In section 2 we describe the sample and the selection criteria and in section 3 the data analysis methods. The observed and rest-frame properties of the GRBs of the sample are discussed in section 4. Our conclusions are presented in section 5. Throughout the paper we assume a standard cosmology with $h={\Omega}_{\Lambda} = 0.7$ and ${\Omega}_{m} = 0.3$. Errors are given at the 68\% confidence level unless stated differently.

\section{The sample: selection and description}

We selected our sample among all the {\it Swift} GRBs with $T_{90} < 2$ s and adding those classified as SGRBs in light of their prompt emission hardness ratio and absence of a spectral lag. With these criteria we could include in our sample also SGRBs whose {\it Swift}-BAT light curve shows a short-duration peak followed by a softer, long-lasting tail (the so-called ``extended emission''). For these events, $T_{90}$ can be longer than 2 s. 
%For GRBs before Dec 2009 we referred to the 2nd {\it Swift}-BAT catalogue (Sakamoto et al. 2011), while for %the others, to the refined analysis GCN circulars\footnote{http://gcn.gsfc.nasa.gov/gcn3.archive.html} of %the {\it Swift}-BAT team. 
We then selected: 1) all the events promptly\footnote{within 120 s from the trigger.} re-pointed by the {\it Swift} X-Ray Telescope (XRT; Burrows et al. 2005) and 2) with favorable observing conditions for ground-based optical follow-up aimed at redshift determination, i.e. events with low Galactic extinction in the direction of the burst ($A_V < 0.5$ mag). Such criteria are inspired, although less tight, to the ones used to built the TOUGH and BAT6 sample for LGRBs (Jakobsson et al. 2006; Hjorth et al. 2012; Salvaterra et al. 2012). In particular we note that, in order to avoid an X-ray selected sample, we do not require a detection of the X-ray afterglow (but just a prompt observation with the {\it Swift}-XRT). 

The sample built this way (the {\it total} sample from now on) consists of 37 SGRBs, with a completeness in redshift of $43\%$ (Table~1). This {\it total} sample includes 5 SGRBs with extended emission (EE) and 19 with short-lived X-ray afterglow (SL)\footnote{Defined by Sakamoto \& Gehrels (2009) as those events for which the X-ray afterglow flux at $10^4\,\mathrm{s}$ after the trigger is less than $10^{-13}\,\mathrm{erg\,cm^{-2}\,s^{-1}}$.}. We restricted this sample to those events with a peak photon flux $P \geq 3.5$ ph s$^{-1}$ cm$^{-2}$, computed using the 15--150 keV {\it Swift-}BAT light curves binned with $\delta t = 64$ ms. This further criterium selects 17 SGRBs, 12 with a measured redshift, providing a sample that has, at the same time, a size large enough to perform statistical studies and a high level of redshift completeness (71\%).
An analogous, although less tight, cut was used in  Salvaterra et al. (2012) to built the BAT6 sample  of LGRBs. Being free of selection effects (except for the flux limit), this sample (the {\it complete} sample from now on), although relatively small, provides a useful benchmark to compare the rest-frame physical properties of SGRB prompt and afterglow emission.

\begin{table*}
\caption{List of GRBs matching the selection criteria of our {\it total} sample described in Sect. 2. GRB belonging to the {\it complete} (flux-limited) sub-sample are marked in boldface. Redshifts are provided in the following references: [1] Leibler \& Berger (2010); [2] Soderberg et al. (2006); [3] Berger et al. 2007; [4] Graham et al. (2009); [5] Berger et al. (2009); [6] D'Avanzo et al. (2009); [7] Leibler \& Berger (2010); [8] Rowlinson et al. (2010); [9] Antonelli et al. (2009); [10] Levesque et al. (2010); [11] McBreen et al. (2010); [12] Fong et al. (2011); [13] Fong et al. (2013); [14] Tanvir et al. (2010); [15] Gorosabel et al. (2010); [16] Chornock \& Berger (2011); [17] Margutti et al. (2012); [18] Sakamoto et al. (2013); [19] Cucchiara et al. (2013).
}
%\centering
%\footnotesize
\begin{tabular}{ccccccccc} \hline 
GRB            &  $T_{90}$      &  $PF_{64}$	        &   EE 	       &   SL/LL &   XRT err	    & OA       &   redshift	& References	  \\
               &    s           &  ph cm$^2$ s$^{-1}$	&  	       &    	 &  $''$	    &  	       &		&	  \\ \hline
050509B        & 0.02           &  $1.3$                &   N          &  SL     &  $3.8$          &  N        &  N$^{a}$	&	  \\
050813         & 0.45           &  $2.1$                &   N          &  SL     &  $2.9$          &  N        &  N		&	  \\
050906         & 0.153          &  $1.7$                &   N          &  $-$      &  $-$            &  N      &  N		&	  \\
051105A        & 0.06           &  $2.7$                &   N          &  $-$      &  $-$            &  N      &  N		&	  \\
051210         & 1.30           &  $1.0$                &   N          &  SL     &  $1.6$          &  N        &  $1.3^{b}$	& 1	  \\
{\bf 051221A}  & 1.40           &  $40.7$               &   N          &  LL     &  $1.4$          &  Y        &  $0.547$	& 2	  \\
051227         & 115.40         &  $1.5$                &   Y          &  LL     &  $3.6$          &  Y        &  N		&	  \\		     
{\bf 060313}   & 0.74           &  $30.9$               &   N          &  LL     &  $1.4$          &  Y        &  N		&	  \\		     
060502B        & 0.14           &  $3.4$                &   N          &  SL     &  $5.2$          &  N        &  N$^{a}$	&	  \\		     
060801         & 0.50           &  $2.1$                &   N          &  SL     &  $1.5$          &  N        &  1.13  	& 3	  \\		     
{\bf 061201}   & 0.78           &  $8.0$                &   N          &  LL     &  $1.4$          &  Y        &  N		&	  \\		     
061217         & 0.24           &  $2.0$                &   N          &  SL     &  $5.5$          &  N        &  N$^{a}$	&	  \\		     
070209         & 0.07           &  $2.8$                &   N          &  $-$      &  $-$            &  N      &  N		&	  \\		     
{\bf 070714B}  & 80.00          &  $8.1$                &   Y          &  LL     &  $1.4$          &  Y        &  0.92  	& 4	  \\		     
070724A        & 0.43           &  $1.5$                &   N          &  LL     &  $1.7$            &  Y      &  0.457 	& 5	  \\		     
070729         & 0.99           &  $1.9$                &   N          &  SL     &  $2.5$          &  N        &  N$^{a}$	&	  \\		     
070809         & 1.28           &  $1.9$                &   N          &  LL     &  $3.6$          &  Y        &  N$^{a}$	&	  \\		     
070810B        & 0.07           &  $2.1$                &   N          &  $-$      &  $-$            &  N      &  N		&	  \\		     
071227         & 144.98         &  $2.9$                &   Y          &  LL     &  $1.7$          &  Y        &  0.381 	& 6	  \\		     
{\bf 080123}   & 115.18         &  $6.1$                &   Y          &  LL     &  $1.7$          &  Y        &  0.495 	& 7	  \\		     
{\bf 080503}   & 159.78         &  $4.0$                &   Y          &  SL     &  $1.6$          &  Y        &  N		&	  \\		     
{\bf 080905A}  & 1.02           &  $3.7$                &   N          &  SL     &  $1.6$          &  Y        &  0.122$^{d}$	& 8	  \\		     
{\bf 090426}$^{c}$   & 1.24           &  $4.7$                &   N          &  LL     &  $1.4$          &  Y  &  2.609 	& 9,10	  \\		     
{\bf 090510}   & 0.30           &  $20.1$               &   N          &  LL     &  $1.4$          &  Y        &  0.903 	& 11	  \\		     
{\bf 090515}   & 0.04           &  $5.2$                &   N          &  SL     &  $2.9$          &  Y        &  N$^{a}$	&	  \\		     
090607$^{c}$         & 2.29           &  $2.0$                &   N          &  SL     &  $3.6$          &  N  &  N		&	  \\		     
{\bf 100117A}  & 0.30           &  $4.4$                &   N          &  SL     &  $3.6$          &  Y        &  0.92  	& 12	  \\		     
{\bf 100625A}  & 0.33           &  $9.3$                &   N          &  SL     &  $1.8$          &  N        &  0.452 	& 13	  \\		     
100816A$^{c}$  & 2.90           &  $12.9$               &   N          &  LL     &  $1.4$          &  Y  &  0.805 	& 14	  \\		     
{\bf 101219A}  & 0.60           &  $8.9$                &   N          &  SL     &  $1.7$          &  N        &  0.718 	& 15,16	  \\		     
101224A        & 0.20           &  $3.1$                &   N          &  SL     &  $3.2$          &  N        &  N		&	  \\		     
110112A        & 0.50           &  $1.1$                &   N          &  SL     &  $1.7$          &  Y        &  N		&	  \\		     
{\bf 111117A}  & 0.47           &  $5.8$                &   N          &  SL     &  $3.6$          &  N        &  $1.3^{b}$	& 17,18	  \\		     
121226A        & 1.00           &  $2.9$                &   N          &  LL     &  $3.5$          &  N        &  N		&	  \\		     
130313A        & 0.26           &  $2.8$                &   N          &  SL     &  $4.8$          &  N        &  N		&	  \\		     
{\bf 130515A}  & 0.29           &  $8.4$                &   N          &  SL     &  $2.3$          &  N        &  N		&	  \\		     
{\bf 130603B}  & 0.18           &  $54.2$               &   N          &  LL     &  $1.4$          &  Y        &  0.356 	& 19	  \\		     
\hline
\hline
\end{tabular}   

\noindent {\bf Notes}. $^a$ GRB with proposed redshift based on the association with the nearby galaxy with the lowest chance probability. GRB\,050509B: $z=0.225$ (Gehrels et al. 2005; Bloom et al. 2006); GRB\,060502B: $z=0.287$ (Bloom et al. 2007); GRB\,061217: $z=0.83$ (Berger et al. 2007); GRB\,070729: $z=0.8$ (Leibler \& Berger 2010); GRB\,070809: $z=0.473$ (Berger 2010); GRB\,090515: $z=0.403$ (Berger 2010).

\noindent $^b$ Photometric redshift.

\noindent $^{c}$ GRB with uncertain classification, possibly long (see Sect. 2.1 and 4.2.2).

\noindent $^{d}$ Possibly underestimated redshift (see Sect. 4.3.1).

\label{tab_log_sample}
\end{table*}

\subsection{GRBs with uncertain classification and actual sample size}

The classification of three GRBs of our sample as SGRB events is uncertain. GRB\,090607 and GRB\,100816A have $T_{90} > 2$ s (see Table~1), however, their spectral hardness and lag are consistent with those of short-hard GRBs (Barthelmy et al. 2009; Norris et al. 2010). GRB\,090426, is definitely a short-duration event with a $T_{90} = 1.2$ s. However, this GRB is consistent at $2\sigma$ with the $E_\mathrm{peak}-E_{\mathrm{iso}}$ correlation (Antonelli et al. 2009), which is followed by LGRBs (Amati et al. 2002), making its classification uncertain (see also Levesque et al. 2010; Th\"one et al. 2012). In the following sections, we will investigate the observer and rest-frame properties of these events, with the aim of shedding light on their classification either as long or short GRBs. As will be shown in Sect 4.2, we propose GRB\,100816A to be a LGRB, in light of its consistency with the $E_\mathrm{peak}-E_\mathrm{iso}$ correlation coupled with its $T_{90} > 2$ s. As a result, our {\it total} and {\it complete} sample consist of 36 and 16 SGRBs, respectively, with a completeness in redshift of 42\% (15/36) and 69\% (11/16), respectively.

\subsection{Short GRBs redshift reliability}

Very recently, the redshift of the exceptionally bright short GRB 130603B has been measured through spectroscopy of its optical afterglow. This is the first clean absorption spectrum obtained for the optical afterglow of a securely-classified SGRB (Cucchiara et al. 2013; de Ugarte Postigo et al. 2013). Optical afterglow spectroscopy of SGRBs have been reported in the past also for GRB\,090426 and GRB\,100816A (Antonelli et al. 2009; Levesque et al. 2010; Tanvir et al. 2010; Gorosabel et al. 2010), whose classification as SGRBs is however highly uncertain (see Sect. 2.1 and 4.2), while with a $T_{90}=0.18$ s, a hard spectrum and negligible spectral lag, GRB 130603B can be classified as a SGRB beyond any doubt (Barthelmy et al. 2013; Norris et al. 2013; Golenetskii et al. 2013). However, apart from such exceptional event, the remaining SGRB redshifts are obtained through spectroscopy of their associated host galaxies. A direct consequence of this is that the SGRB-host galaxy association can only be secured when the optical afterglow is detected and found to lie within the host galaxy light with a sub-arcsecond precision or proposed on chance probability arguments (and not, e.g., by matching the redshift measured through spectroscopy of both the optical afterglow and the host galaxy). Throughout the paper, we will consider as GRBs with a secure redshift measurement only those events for which an optical afterglow was found to lie within the host galaxy light or those events having a host galaxy whose position is within a precise (radius$<2''$) X-ray error circle. 

These tight requirements we have put on the redshift reliability can bias our sample against events with a large offset with respect to their host galaxy. However, we note that with such criteria we exclude just one event whose redshift is proposed on the basis of probabilistic association with a nearby host galaxy (namely, GRB\,090515; see Table~1).

\section{Data analysis}

\subsection{Observer-frame properties}

In Table~\ref{tab_fluence_flux} we report the prompt emission fluence observed in the 15-150 keV {\it Swift}-BAT energy range, together with the 0.3-10 keV X-ray fluence observed with {\it Swift}-XRT.
We collected the $T_{90}$ and gamma-ray fluence values from the 2nd {\it Swift}-BAT catalogue (Sakamoto et al. 2011) for GRBs before Dec 2009, while for events occuring later than this date, we referred to the refined analysis GCN circulars\footnote{http://gcn.gsfc.nasa.gov/gcn3.archive.html} of the {\it Swift}-BAT team. 
For each GRB of our {\it total} sample we retrieved the 0.3-10 keV flux calibrated X-ray light curves from the automated data products provided by the {\it Swift} Burst analyzer\footnote{http://http://www.swift.ac.uk/burst\_analyser/} (Evans et al. 2009). These light curves were fitted with power laws, broken power laws or multiply broken power laws. 
We then computed the early X-ray fluences observed in the first 500 s of the {\it Swift}-XRT observations and the total X-ray fluences, obtained integrating the observed 0.3-10 keV flux under the best light curve fit between 120s and 620 s after the BAT trigger\footnote{This choice is motivated by the will of having the earliest observation time common to all the burst of the sample and to integrate over a time long enough to measure the early X-ray fluences with acceptable ($\sim 3\sigma$) significance.} and under the whole light curve (from the first to the last X-ray afterglow detection), respectively. For GRBs with just one detection of the X-ray afterglow (seven events, see Table~2) it was not possible to constrain a decay and were not considered in this analysis. 

\begin{table*}
\caption{High energy properties of the GRB of our {\it total} sample in the observer frame: prompt emission fluence measured by the {\it Swift-}BAT in the 15-150 keV energy range and X-ray fluence measured by the {\it Swift-}XRT in the 0.3-10 keV energy range. The early and total X-ray fluences are measured integrating the observed 0.3-10 keV flux under the best light curve fit between 120 s and 620 s after the BAT trigger and under the whole light curve (from the first to the last X-ray afterglow detection), respectively (see Sect. 3.1 for details). GRB belonging to the {\it complete} (flux-limited) sub-sample are marked in boldface.
}
%\centering
%\footnotesize
\begin{tabular}{cccc} \hline 
GRB            &  $\gamma$ prompt fluence    &  X-ray early fluence    &   X-ray total fluence 	           \\
                  & $10^{-7}$ erg cm$^2$     & $10^{-8}$ erg cm$^2$    & $10^{-8}$ erg cm$^2$              \\ \hline
050509B        & $ 0.07 \pm 0.02$         &  $0.22 \pm 0.07$	       & $0.42 \pm 0.11 $ 	     \\
050813         & $ 0.37 \pm 0.11$         &  $0.08 \pm 0.02$	       & $-^b$	     \\
050906         & $ 0.06 \pm 0.02$         &  $-^a$		       & $-^a$	                     \\
051105A        & $ 0.17 \pm 0.04$         &  $-^a$		       & $-^a$	                     \\
051210         & $ 0.86 \pm 0.14$         &  $5.63 \pm 1.16$	       & $7.74 \pm 1.93 $	     \\
{\bf 051221A}  & $11.60 \pm 0.35$         &  $4.72 \pm 0.96$	       & $20.15 \pm 5.04 $	     \\
051227         & $ 7.15 \pm 1.10$         &  $5.46 \pm 1.10$	       & $10.04 \pm 2.51 $	     \\ 	      
{\bf 060313}   & $11.40 \pm 0.46$	  &  $7.83 \pm 1.73$	       & $27.15 \pm 6.79 $	     \\ 		
060502B        & $ 0.49 \pm 0.07$	  &  $0.13 \pm 0.03$	       & $-^b$	     \\ 		
060801         & $ 0.81 \pm 0.10$	  &  $2.55 \pm 0.56$	       & $3.15 \pm 0.79 $	     \\ 		
{\bf 061201}   & $ 3.41 \pm 0.28$	  &  $12.34 \pm 2.70$	       & $47.06 \pm 11.76$	     \\ 		
061217         & $ 0.43 \pm 0.08$	  &  $0.10  \pm 0.07$	       & $-^b$	                     \\ 		
070209         & $ 0.22 \pm 0.04$	  &  $-^a$		       & $-^a$	                     \\ 		
{\bf 070714B}  & $ 7.30 \pm 1.04$	  &  $12.88 \pm 2.39$	       & $34.25 \pm 8.56 $	     \\ 		
070724A        & $ 0.31 \pm 0.07$	  &  $3.81 \pm 0.83$	       & $7.01 \pm 1.75 $	     \\ 		
070729         & $ 1.03 \pm 0.17$	  &  $0.22 \pm 0.04$	       & $-^b$	     \\ 		
070809         & $ 1.02 \pm 0.15$	  &  $0.33 \pm 0.08$	       & $4.99 \pm 1.25 $	     \\ 		
070810B        & $ 0.16 \pm 0.04$	  &  $-^a$		       & $-^a$	                     \\ 		
071227         & $ 5.05 \pm 1.23$	  &  $4.00 \pm 0.72$	       & $7.52 \pm 1.88 $	     \\ 		
{\bf 080123}   & $ 5.70 \pm 1.70$	  &  $16.09 \pm 2.60$	       & $19.58 \pm 4.90 $	     \\ 		
{\bf 080503}   & $20.00 \pm 1.49$	  &  $25.31 \pm 3.99$	       & $46.02 \pm 11.51$	     \\ 		
{\bf 080905A}  & $ 1.40 \pm 0.19$	  &  $4.99 \pm 1.09$	       & $5.59 \pm 1.40 $	     \\ 		
{\bf 090426}   & $ 1.85 \pm 0.27$	  &  $1.55 \pm 0.35$	       & $9.38 \pm 2.34 $	     \\ 		
{\bf 090510}   & $ 3.40 \pm 0.40$	  &  $14.40 \pm 2.97$	       & $37.04 \pm 9.26 $	     \\ 		
{\bf 090515}   & $ 0.22 \pm 0.04$	  &  $7.35 \pm 1.12$	       & $4.22 \pm 1.06 $	     \\ 		
090607         & $ 1.14 \pm 0.19$	  &  $4.36 \pm 0.85$	       & $1.57 \pm 0.39 $	     \\ 		
{\bf 100117A}  & $ 0.93 \pm 0.13$	  &  $10.84 \pm 2.71$	       & $18.66 \pm 4.67$	     \\ 		
{\bf 100625A}  & $ 2.30 \pm 0.20$	  &  $0.98 \pm 0.22$	       & $1.62 \pm 0.40 $	     \\ 		
{\bf 100816A}  & $20.00 \pm 1.00$	  &  $7.04 \pm 1.36$	       & $23.12 \pm 5.78 $	     \\ 		
{\bf 101219A}  & $ 4.60 \pm 0.30$	  &  $4.54 \pm 1.00$	       & $4.88 \pm 1.22 $	     \\ 		
101224A        & $ 0.58 \pm 0.11$	  &  $0.44 \pm 0.21$	       & $-^b$	     \\ 		
110112A        & $ 0.30 \pm 0.09$	  &  $0.48 \pm 0.11$	       & $1.72 \pm 0.43 $	     \\ 		
{\bf 111117A}  & $ 1.40 \pm 0.18$	  &  $0.95 \pm 0.21$	       & $1.84 \pm 0.46 $	     \\ 		
121226A        & $ 1.40 \pm 0.20$	  &  $3.77 \pm 0.84$	       & $16.58 \pm 4.14 $	     \\ 		
130313A        & $ 0.31 \pm 0.07$	  &  $0.66 \pm 0.31$	       & $-^b$	     \\ 		
{\bf 130515A}  & $ 1.50 \pm 0.20$	  &  $0.19 \pm 0.03$	       & $-^b$	     \\ 		
{\bf 130603B}  & $ 6.30 \pm 0.30$	  &  $6.48 \pm 1.42$	       & $42.92 \pm 10.73$	     \\ 		
\hline
\hline
\end{tabular}   

\noindent {\bf Notes}. $^a$ GRB without an XRT detected afterglow.

\noindent $^b$ GRB with just one detection of the X-ray afterglow.

\label{tab_fluence_flux}
\end{table*}

\subsection{Rest-frame properties}

\subsubsection{$\gamma-$ and X$-$rays spectral analysis}

To collect the prompt emission spectral properties of the bursts in our {\it complete} sample we refer to the spectral analysis reported in the GCN circulars or to the refined analysis reported in published papers, when available. Most of the GRBs in our sample have been detected not only by {\it Swift}-BAT, but also by other high energy satellites, including {\it Konus}-WIND, {\it Fermi}-GBM and {\it Suzaku}-WAM, providing a complete information on the prompt spectrum (peak energy, fluence and peak flux). For the bursts with a measured redshift it was possible to estimate the rest frame $E_\mathrm{peak}$, $E_{\mathrm{iso}}$, and $L_{\mathrm{iso}}$ and test the $E_\mathrm{peak}-E_{\mathrm{iso}}$ and $E_\mathrm{peak}-L_{\mathrm{iso}}$ correlations. $E_{\mathrm{iso}}$ and $L_{\mathrm{iso}}$ have been estimated in the (rest frame) energy range 1 keV -- 10 MeV (Table~\ref{tab_spettro_prompt}). 
%For this burst the peak energy has been derived from the spectral analysis by fixing the value of the photon %indices to $\alpha=-1$ and $\beta=-2.3$ (Antonelli et al. 2009).
%Note that the sample of events with measured redshift includes GRB\,090426 and GRB\,100816A whose %classification as SGRB is uncertain (see Sect. 2.1).

We retrieved the automated data products provided by the \textit{Swift}-XRT GRB spectrum repository\footnote{http://www.swift.ac.uk/xrt\_spectra/.} to obtain the intrinsic X-ray absorbing column densities for the GRBs of our {\it complete} sample with measured redshift. Following the procedure described in Kopac et al. (2012) we used data mostly from photon counting (PC) mode in the widest time epoch where the $0.3-1.5\,\mathrm{keV}$ to $1.5-10\,\mathrm{keV}$ hardness ratio is constant, in order to prevent spectral changes that can affect the X-ray column density determination. Each spectra is fitted with the {\it phabs*zphabs*pow} model within the XSPEC package. The first absorbtion component ({\it phabs}) is frozen at the Galactic contribution to the X-ray $N_H$ in the direction of each GRB (using the value computed by Kalberla et al. 2005). The second (intrinsic) absorption component ({\it zphabs}) is left free to vary, with the redshift frozen to the value reported in the literature for any given GRB (Table~1). The third component ({\it pow}) is a simple power-law model with photon index and normalization free to vary. The results of our analysis are shown in Table~4.

\begin{table*}
\caption{
Prompt emission rest frame spectral properties for the SGRBs of the {\it complete} sample with measured redshift. Spectral parameters, rest frame peak energies ($E_\mathrm{peak}$), isotropic energies
($E_{\mathrm{iso}}$) and luminosities ($L_{\mathrm{iso}}$) are listed. Columns 7 and 8 report the name of the mission from which the spectral properties have been derived (F={\it Fermi}, 
K={\it Konus/Wind},Su={\it Suzaku}, S={\it Swift}) and the
width of the temporal bin chosen to rebin the prompt emission light curve ($\Delta t$), respectively. Reference in the last column are for the spectral properties:
[1] Golenetskii et al. (2005), [2] Ohno et al. (2007), [3] Uehara et al. (2008), [4] Nava et al. (2011), [5] Antonelli et al. (2009), [6] Bhat et al. (2010), [7]
Fitzpatrick (2010), [8] Golenetskii et al. (2010), [9] Sakamoto et al. (2013), [10] Golenetskii et al. (2013).
}
\centering
%\footnotesize
\begin{tabular}{ccccccccc} \hline 
GRB     &  $z$    & $\alpha \, [\beta]$   &  $E_\mathrm{peak}$		       &  $E_{\mathrm{iso}}$	      &  $L_{\mathrm{iso}}$ 	                    & Mission     &	$\Delta t$  & Ref.   \\
 &          &        	                  &  keV		       &  $10^{51}$ erg	      &  $10^{52}$ erg s$^{-1}$             &       &   ms	   &   \\ \hline
051221A & 0.547  & $-1.08 \pm 0.13$      &  $ 621.49 \pm  126.77$     &  $2.63  \pm 0.33 $   &  $5.84  \pm 0.89 $                  & K     &   $    4$      & 1   \\
070714B & 0.92    & $-0.86 \pm 0.10$      &  $2150.40 \pm 1044.48$     &  $9.80  \pm 2.38 $   &  $1.30  \pm 0.14 $                  & Su    &   $ 1000$      & 2  \\
080123  & 0.495   & $-1.20 \pm 0.38$      &  $ 104.65^a$	       &  $0.13^{a}$	      &  $0.017^{a}$	                    & Su    &   $  880$      & 3  \\
080905A & 0.122  & $-0.13 \pm 0.16$      &  $ 578.95 \pm   77.42$     &  $0.032 \pm 0.003$   &  $0.012 \pm 0.002$                  & F     &   $   64$      & 4  \\
090426  & 2.609   & $-1 \, [-2.3]$        &  $ 176.84 \pm   72.18^b$   &  $5.41  \pm 0.64^b$  &  $1.76  \pm 0.22^b$                 & S     &   $ 1000$      & 5  \\
090510  & 0.903   & $-0.82 \pm 0.02 \, [-2.76 \pm 0.25]$   &  $8089.65 \pm  593.74$     &  $74.30 \pm 3.21 $   &  $17.80 \pm 1.17 $ & F     &   $   64$  & 4  \\
100117A & 0.920   & $-0.15 \pm 0.21$      &  $ 549.12 \pm   84.48$     &  $0.81  \pm 0.10 $   &  $0.97  \pm 0.13 $                  & F     &   $   64$      & 4  \\
100625A & 0.452   & $-0.60 \pm 0.11$      &  $ 701.32 \pm  114.71$     &  $0.75  \pm 0.03 $   &  $0.34  \pm 0.01 $                  & F     &   $   64$      & 6  \\ 
100816A & 0.805   & $-0.31 \pm 0.05 \, [-2.77 \pm 0.17]$   &  $ 247.29 \pm    8.48$     &  $7.27  \pm 0.25 $   &  $0.72  \pm 0.01 $ & F     &   $ 1024$  & 7  \\ 
101219A & 0.718   & $-0.22 \pm 0.27$      &  $ 841.82 \pm  154.62$     &  $4.87  \pm 0.68 $   &  $6.50  \pm 1.86 $                  & K     &   $   16$      & 8  \\
111117A & 1.3$^c$ & $-0.28 \pm 0.28$      &  $ 966.00 \pm  322.00$     &  $3.38  \pm 1.06 $   &  $4.04  \pm 1.28 $                  & F     &   $   50$      & 9  \\ 
130603B & 0.356   & $-0.73 \pm 0.15$      &  $ 894.96 \pm  135.60$     &  $2.12  \pm 0.23 $   &  $4.35  \pm 0.87 $                  & K     &   $   16$      & 10  \\
\hline
\label{tab_spettro_prompt}
\end{tabular}   

\noindent {\bf Notes}. $^a$ Lower limit.

\noindent $^b$ Values obtained from fitting with a Band function with $\alpha=-1$ and $\beta=-2.3$ fixed.

\noindent $^c$ Photometric redshift.

\label{tab_Eiso}
\end{table*}

\begin{table*}
\begin{center}
\caption{Intrinsic X-ray absorbing column densities of the SGRB of our {\it complete} sample with measured redshift. Columns are: GRB indentifier, redshift, Galactic X-ray absorbing column density, X-ray spectrum
exposure time and time interval, spectral photon index, intrinsic X-ray absorbing column desity. Errors are at $90\%$ confidence level.}
\begin{tabular}{ccccccccccccc}
\hline
GRB      & $z$      & $\mathrm{N_H}$ Gal.          & Exp. (interval)                 & $\Gamma$               & $\mathrm{N_H}(z)$            \\
         &          & $10^{20}\,\mathrm{cm^{-2}}$  &     ks (s)                      &                        & $10^{21}\,\mathrm{cm^{-2}}$  \\
\hline
051221A  & $0.547$  & $5.7$                        & $68.9$ ($6000-2 {\times} 10^5$) & $2.02_{-0.16}^{+0.17}$ & $2.0_{-0.8}^{+0.9}$          \\
070714B  & $0.923$  & $6.4$                        & $25.9$ ($450-6 {\times} 10^4$)  & $1.90_{-0.16}^{+0.17}$ & $2.1_{-1.2}^{+1.4}$          \\
080123   & $0.495$  & $2.3$                        & $2.2$ ($227-2459$)              & $2.15_{-0.30}^{+0.39}$ & $< 3.1^\mathrm{UL}$          \\
080905A  & $0.122$  & $9.0$                        & $0.9$ ($330-1230$)              & $1.70_{-0.50}^{+0.57}$ & $< 5.7^\mathrm{UL}$          \\
090426   & $2.609$  & $1.5$                        & $6.8$ ($100-2.5 {\times} 10^4$) & $2.05_{-0.14}^{+0.15}$ & $< 8.8^\mathrm{UL}$          \\
090510   & $0.903$  & $1.7$                        & $0.14$ ($100-240$)              & $1.76_{-0.15}^{+0.16}$ & $1.8_{-1.0}^{+1.2}$          \\
100117A  & $0.915$  & $2.7$                        & $0.05$ ($105-155$)              & $1.50_{-0.22}^{+0.26}$ & $3.6_{-2.5}^{+4.1}$          \\
100625A  & $0.452$  & $2.2$                        & $0.65$ ($61-708$)               & $2.34_{-0.22}^{+0.28}$ & $< 1.8^\mathrm{UL}$          \\
100816A  & $0.805$  & $4.5$                        & $50.9$ ($201-3 {\times} 10^5$)  & $1.99_{-0.15}^{+0.16}$ & $3.0_{-1.2}^{+1.3}$          \\
101219A  & $0.718$  & $4.9$                        & $0.8$  ($64-898$)               & $1.50_{-0.23}^{+0.25}$ & $4.5_{-2.4}^{+3.1}$          \\
111117A  & $1.3^a$    & $3.7$                        & $0.58$ ($89-668$)               & $1.84_{-0.32}^{+0.36}$ & $6.2_{-4.7}^{+6.8}$          \\
130603B  & $0.356$  & $1.9$                        & $15.8$ ($104-1 {\times} 10^5$)  & $2.07_{-0.14}^{+0.14}$ & $3.2_{-0.7}^{+0.8}$          \\
\hline
\end{tabular}
\end{center}
\noindent $^a$ Photometric redshift.
\label{tab:N_H}
\end{table*}

\subsubsection{X-ray luminosities}

Using the automated data products provided by the {\it Swift} Burst analyzer (Evans et al. 2009) we estimated the afterglow X-ray integral fluxes in the 2-10 keV rest frame common energy band and computed the corresponding rest frame X-ray luminosities at different rest frame times for all the GRBs of our {\it complete} sample with a measured redshift. The 2-10 keV rest frame fluxes were computed from the observed  0.3-10 keV unabsorbed fluxes and the time-resolved measured photon spectral index, $\Gamma$, (that we retrieved from the {\it Swift} Burst Analyzer) in the following way (see also Gehrels et al. 2008):  

\begin{equation} 
f_{X,rf}(2-10 \, {\rm{keV}}) = f_X(0.3-10 \, \rm{keV})\frac{\left({\frac{10}{1+z}}\right)^{2-\Gamma}-\left({\frac{2}{1+z}}\right)^{2-\Gamma}}{{10}^{2-\Gamma}-{0.3}^{2-\Gamma}}
\label{kcorr_eq}
\end{equation}

\noindent The best-fit of each GRB X-ray light curve was interpolated or extrapolated to the given rest frame times. These afterglow X-ray luminosities were compared with the prompt emission isotropic energies $E_{\mathrm{iso}}$, the isotropic peak luminosities $L_{\mathrm{iso}}$ and the rest frame peak energies $E_\mathrm{peak}$ for the bursts of our {\it complete} sample. The obtained 2-10 keV rest frame X-ray luminosities are reported in Table~\ref{tab_Xlum}.

\begin{table*}
\caption{X-ray luminosities in the 2--10 keV rest frame energy band computed at four different rest frame times (5 min, 1 hr, 11 hr and 24 hr) for the SGRBs of the {\it complete} sample with measured redshift (see Sect. 2 for details). 
}
\centering
%\footnotesize
\begin{tabular}{cccccccccc} \hline 
GRB     &  $z$      &  $L_{X,5}$	      &   $L_{X,5}$ err 	&   $L_{X,1}$		  &   $L_{X,1}$ err	    & $L_{X,11}$	      &   $L_{X,11}$ err 	& $L_{X,24}$		  &   $L_{X,24}$ err	       \\
        &           &  erg s$^{-1}$	      &  erg s$^{-1}$		&  erg s$^{-1}$ 	  &  erg s$^{-1}$	    &  erg s$^{-1}$	      &  erg s$^{-1}$		&  erg s$^{-1}$ 	  &  erg s$^{-1}$	       \\ \hline
051221A & 0.547     &  $1.65 \times 10^{46}$ &  $3.31 \times 10^{45}$ &  $1.40 \times 10^{45}$ &  $3.50 \times 10^{44}$ &  $4.87 \times 10^{44}$ &  $1.71 \times 10^{44}$ &  $8.66 \times 10^{43}$ &  $3.46 \times 10^{43}$	\\
070714B & 0.92       &  $9.86 \times 10^{46}$ &  $1.97 \times 10^{46}$ &  $1.04 \times 10^{46}$ &  $2.59 \times 10^{45}$ &  $3.99 \times 10^{43}$ &  $1.39 \times 10^{43}$ &  $9.19 \times 10^{42}$ &  $3.68 \times 10^{42}$	\\
080123  & 0.495      &  $7.82 \times 10^{44}$ &  $1.56 \times 10^{44}$ &  $5.32 \times 10^{43}$ &  $1.33 \times 10^{43}$ &  $3.62 \times 10^{43}$ &  $1.27 \times 10^{43}$ &  $3.19 \times 10^{43}$ &  $1.28 \times 10^{43}$	\\
080905  & 0.122     &  $9.02 \times 10^{44}$ &  $1.80 \times 10^{44}$ &  $3.85 \times 10^{43}$ &  $4.35 \times 10^{41}$ &  $4.18 \times 10^{39}$ &  $1.46 \times 10^{39}$ &  $5.88 \times 10^{38}$ &  $2.35 \times 10^{38}$	\\
090426  & 2.609      &  $2.09 \times 10^{47}$ &  $4.18 \times 10^{46}$ &  $2.00 \times 10^{46}$ &  $5.00 \times 10^{45}$ &  $2.08 \times 10^{45}$ &  $7.29 \times 10^{44}$ &  $9.97 \times 10^{44}$ &  $3.99 \times 10^{44}$	\\
090510  & 0.903      &  $3.32 \times 10^{47}$ &  $6.64 \times 10^{46}$ &  $6.65 \times 10^{45}$ &  $1.66 \times 10^{45}$ &  $4.43 \times 10^{43}$ &  $1.55 \times 10^{43}$ &  $8.67 \times 10^{42}$ &  $3.47 \times 10^{42}$	\\
100117A & 0.920      &  $1.73 \times 10^{46}$ &  $3.47 \times 10^{45}$ &  $9.96 \times 10^{43}$ &  $2.49 \times 10^{43}$ &  $6.86 \times 10^{41}$ &  $2.40 \times 10^{41}$ &  $1.36 \times 10^{41}$ &  $5.43 \times 10^{40}$	\\ 
100625A & 0.452      &  $1.53 \times 10^{45}$ &  $3.05 \times 10^{44}$ &  $3.22 \times 10^{43}$ &  $8.04 \times 10^{42}$ &  $7.75 \times 10^{41}$ &  $2.71 \times 10^{41}$ &  $2.32 \times 10^{41}$ &  $9.23 \times 10^{40}$	\\ 
100816A & 0.805      &  $5.88 \times 10^{46}$ &  $1.18 \times 10^{46}$ &  $3.83 \times 10^{45}$ &  $9.57 \times 10^{44}$ &  $3.27 \times 10^{44}$ &  $1.15 \times 10^{44}$ &  $1.47 \times 10^{44}$ &  $5.88 \times 10^{43}$	\\ 
101219A & 0.718      &  $1.08 \times 10^{46}$ &  $2.16 \times 10^{45}$ &  $3.76 \times 10^{43}$ &  $9.41 \times 10^{42}$ &  $1.60 \times 10^{41}$ &  $5.59 \times 10^{40}$ &  $2.70 \times 10^{40}$ &  $1.08 \times 10^{40}$	\\
111117A & 1.3$^a$    &  $2.17 \times 10^{46}$ &  $4.35 \times 10^{45}$ &  $7.17 \times 10^{44}$ &  $1.79 \times 10^{44}$ &  $2.67 \times 10^{43}$ &  $9.33 \times 10^{42}$ &  $9.13 \times 10^{42}$ &  $3.65 \times 10^{42}$	\\ 
130603B & 0.356      &  $1.68 \times 10^{46}$ &  $3.36 \times 10^{45}$ &  $3.29 \times 10^{45}$ &  $8.21 \times 10^{44}$ &  $6.81 \times 10^{43}$ &  $2.38 \times 10^{43}$ &  $1.93 \times 10^{43}$ &  $7.71 \times 10^{42}$	\\
\hline						   
\end{tabular}   

\noindent $^a$ Photometric redshift.

\label{tab_Xlum}
\end{table*}

\subsection{Correlation Analysis}

In the following sections, we will show the results obtained by testing the existence of correlations among the observed and rest-frame physical properties of the SGRBs of our sample. We computed for each case the Spearman rank correlation coefficient $r$ (Spearman 1904; Press et al. 1986) and, in order to determine the significance of the correlation, the associated null-hypothesis probability $P_{null}$. Given the large scatter of the data points (larger than the uncertainties on each value) and that, in principle, any of the quantities at play can be assumed as the independent variable, we fit the data using the ordinary least squares bisector method (Isobe et al. 1990). 

We note that, for the events of our {\it complete} sample, the rest-frame properties like luminosities and energies can be correlated with the redshift and this could give rise to spurious correlations. Indeed, for any flux-limited sample (as in our case) there will be an inevitable and tight correlation between luminosity and redshift.
This arises because, for a fixed flux limit, only the more powerful sources can be detected out to great distances (see, e.g., Blundell, Rawlings \& Willot 1999 and references therein). In order to properly handle this problem the correlations between luminosities or between luminosity and energy should be examined taking into account the common redshift dependence. This can be done with a partial correlation analysis. If $r_{ij}$ is the correlation coefficient between $x_i$ and $x_j$, in the case of three variables the correlation coefficient between two of them, excluding the effect of the third one is:

\begin{equation} 
r_{12,3} = \frac{r_{12}-r_{13}r_{23}}{\sqrt{1-r^{2}_{13}}\sqrt{1-r^{2}_{23}}}
\label{pad_eq}
\end{equation}

\noindent (Kendall \& Stuart 1979; see also Padovani 1992) where, for our study, the coefficients 1 and 2 refer to the luminosity and energy, respectively, and the coefficient 3 to the redshift. We also performed a linear fit to each data distribution in logarithmic space. The results of such correlation analysis are reported in Table~\ref{tab:log_corr_obs} and Table~\ref{tab:log_corr_rf}.

\section{Results and discussion}

\subsection{{\it Total} sample: properties in the observer frame}

\begin{figure}
   \centering
   \includegraphics[angle=-90,width=84mm]{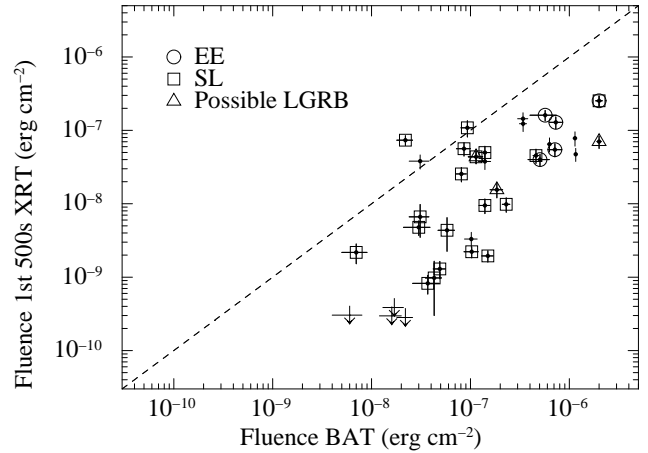}
  \caption{X-ray (0.3--10 keV) fluence during the first 500 s of {\it Swift}-XRT observation {\it vs.} gamma--ray (15-150 keV) prompt fluence of the SGRBs of the {\it total} sample (dots). The dashed line shows where the two quantities are equal (even if in different energy bands). Sample sub-classes are also shown: no XRT prompt detection (lower limits); extended emission SGRBs (open circles); short lived SGRBs (open squares); possible LGRBs (open triangles).
}
       \label{fig:1}
\end{figure}

As discussed in Sect.~1, SGRB progenitors (binary systems of compact objects) can originate from the evolution of massive stars in a primordial binary or by dynamical interactions and capture in globular clusters during their core collapse. In primordial systems, the delay between binary formation and merging is driven by the gravitational wave inspiral time, which is strongly dependent on the initial system separation. Some systems are thus expected to drift away from the star--forming regions in which they formed, before merging takes place, also because they experience a natal kick at the time of the formation of the compact object. Simulations (Belczynski, Bulik \& Kalogera 2002; Belczynski et al. 2006) show that a large fraction of the merging events should take place in the outskirts or even outside the galaxies, in low density environments. A low density circumburst environment is expected also for short GRBs of dynamical origin occurring in globular clusters. For these events, the resulting time delay between star-formation and merging would be dominated by the cluster core-collapse time and thus be comparable to the Hubble time (Hopman et al. 2006). A much faster evolutionary channel has been proposed (Belczynski \& Kalogera 2001; Perna \& Belczynski 2002; Belczynski et al. 2006), leading to merging in only $\sim 10^6-10^7\,\mathrm{yr}$, when most systems are still immersed in their star-forming regions. According to the above scenario, with the exception of the events originated by the ``fast'' primordial channel, SGRBs are generally expected to occur in regions where the density of the diffuse medium is low, giving rise to fainter afterglows, setting in at later times than those of long GRBs (e.g. Vietri 2000; Panaitescu, Kumar \& Narayan 2001; Salvaterra et al. 2010).

\begin{figure}
   \centering
   \includegraphics[angle=-90,width=84mm]{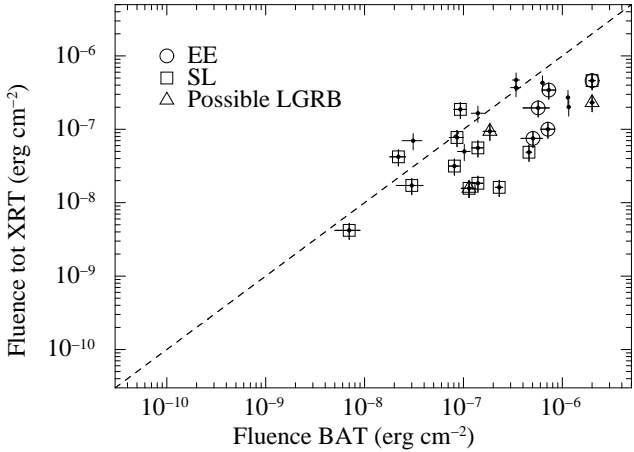}
  \caption{X-ray (0.3--10 keV) total fluence {\it vs.} gamma--ray (15-150 keV) prompt fluence of the SGRBs of the {\it total} sample. The dashed line shows where the two quantities are equal (even if in different energy bands). Sample sub-classes are also shown: extended emission SGRBs (open circles); short lived SGRBs (open squares); possible LGRBs (open triangles).
}
       \label{fig:2}
\end{figure}

Four events (i.e., $13\%$ of the {\it total} sample) have no X-ray afterglow detected, in spite of the prompt {\it Swift} XRT follow-up. This marks a significant difference with respect to the LGRBs, where only $\sim 2\%$ of the events promptly observed by the {\it Swift} XRT is missing an X-ray afterglow detection (Evans et al. 2009). The lack of detection of an X-ray afterglow for these four events can have multiple explanations. It can be ascribed to a difference in the GRB energetics (they could be sub-energetic GRBs), to the low density of the circumburst medium or to a high-redshift origin. For each GRB of the {\it total} sample we computed the observed 0.3-10 keV X-ray fluence during the first 500 s of the {\it Swift}-XRT observation (between 120 s and 620 s after the burst; Table~\ref{tab_fluence_flux}). We then compared it with the prompt emission fluence computed in the 15-150 keV {\it Swift}-BAT energy band (Table~\ref{tab_fluence_flux}). As can be seen in Fig.~1 and Table~\ref{tab:log_corr_obs}, the two quantities scale linearly, with the highest X-ray fluences corresponding to the highest gamma-ray fluences. The four X-ray fluence upper limits lay in the faint end of the BAT fluence distribution, suggesting that these events were probably less energetic, rather than occurred in low-density environments. A similar explanation was proposed also by Nysewander et al. (2009) for SGRBs with faint X-ray afterglows. However, if we assume that the early time X-ray emission is the tail of the prompt emission (i.e. originated by internal shocks; Kumar \& Panaitescu 2000; Tagliaferri et al. 2005), then its intensity should be unrelated to the density of the circumburst medium and clearly tightly correlated to the energetic of the prompt emission. Actually, as we will show in the next sections, the early time X-ray luminosity for the SGRBs of the {\it complete} sample is found to strongly correlate with the isotropic energy of the prompt emission. A useful test to check for the role of the density of the circumburst medium could come from the emission in the optical band (if any), which is expected to be pure external shock afterglow emission even at early times. Unfortunately, none of these SGRBs without an early X-ray detection have an optical afterglow (Berger 2014). Finally, we note that a high redshift origin for these events (all missing a redshift measurement) cannot be ruled out anyway (see, e.g., Berger et al. 2007).

\begin{table*}
   \centering
\caption{Correlation fits and coefficients for the observer frame properties of the {\it total} sample. Data distributions shown in Fig.~1 and Fig.~2 were fitted with the function $y = 10^Ax^B$ (see Sect. 2.1 for details).}
\begin{tabular}{cccccc}
\hline
Correlation	 & A		     & B	       & $r$  & $P_{null}$	       & Dispersion \\
                 & 		     & 		       &      & 		       &	    \\ \hline
X$-$ray$_\mathrm{early}$ $vs.$ $\gamma-$ray   &  $-0.09 \pm 1.02$ & $1.12 \pm 0.11$ & 0.65 & $5.66 \times 10^{-5}$  & 0.384   \\
X$-$ray$_\mathrm{total}$ $vs.$ $\gamma-$ray   &  $-1.21 \pm 0.74$ & $0.88 \pm 0.07$ & 0.74 & $1.55 \times 10^{-5}$  & 0.291   \\ \hline
\hline
\end{tabular}
\label{tab:log_corr_obs}
\end{table*}

We also compared the {\it Swift}-BAT fluence with the total X-ray fluence, computed by integrating the 0.3-10 keV X-ray flux under each light curve best fit (see Sect. 3.1). Again the two quantities follow a broadly linear correlation, suggesting that the X-ray afterglow brightness is a proxy of the prompt emission energy (Fig.~2, Table~\ref{tab:log_corr_obs}). Such results are similar to what found for the observed gamma and X-ray properties of LGRBs (Gehrels et al. 2008; Nysewander et al. 2009; Margutti et al. 2013). In particular, we find that SGRBs with extended emission and SGRBs with short lived X-ray afterglows follow the correlation as the other events of the sample, with the former having sistematically larger $\gamma-$ray fluences, as expected given their longer duration in terms of $T_{90}$. According to these findings, the prompt emission or the X-ray afterglow duration do not seem to provide a unique indicator of a specific progenitor and/or environment for SGRBs.

\subsection{{\it Complete} sample: properties in the rest frame}

\subsubsection{SGRBs spectral hardness}

One of the key properties characterizing the SGRBs is their prompt emission spectrum, which is found to be typically harder with respect to LGRBs (Kouveliotou et al. 1993). Considering the GRB prompt emission spectrum as described by a Band function (Band et al. 1993),  the SGRB spectral hardness is found to be due to a combination of an harder low-energy spectral component (the $\alpha$ index of the Band function) and to a higher spectral peak energy (Ghirlanda et al. 2009). However, these differencies become less significant when the analysis is restricted to the first 1-2 s of the LGRBs prompt emission (Ghirlanda, Ghisellini \& Celotti 2004; Ghirlanda et al. 2009). At the same time, Ghirlanda, Ghisellini \& Celotti (2004) showed that for the brightest SGRBs, the
difference in the spectral hardness with respect to LGRBs is mainly driven by a harder low energy spectral index present in short bursts, rather than due to a different peak energy.
To further investigate this issue, we compared the low-energy spectral indices ($\alpha$) and the peak energies ($E_\mathrm{peak}$) of the SGRBs of our {\it complete} sample (Table~\ref{tab_spettro_prompt}) with the equivalent values of the LGRBs of the BAT6 sample (Nava et al. 2012). 
The obtained distributions have average $\alpha_{SGRB} = -0.6 \, (\sigma=0.4)$ versus $\alpha_{LGRB} = -1.0 \, (\sigma=0.2)$ and $E_\mathrm{peak}^{SGRB} = 1557 \, (\sigma=2352)$ keV versus $E_\mathrm{peak}^{LGRB} = 737 \, (\sigma=510)$ keV. A Kolmogorv-Smirnov (K-S) test gives a 1\% and 29\% probability for the two distributions of $\alpha$ and $E_\mathrm{peak}$ to be drawn from the same population, respectively. This provides a moderate indication (although both distributions are rather scattered) that for a sub-sample of bright events, the spectral hardness of SGRBs (compared to LGRBs) is mainly due to a different low-energy spectral index (in agreement with the findings of Ghirlanda, Ghisellini \& Celotti 2004).

\subsubsection{Prompt spectral energy correlations}

\begin{figure*}
   \centering
   \includegraphics[width=162mm]{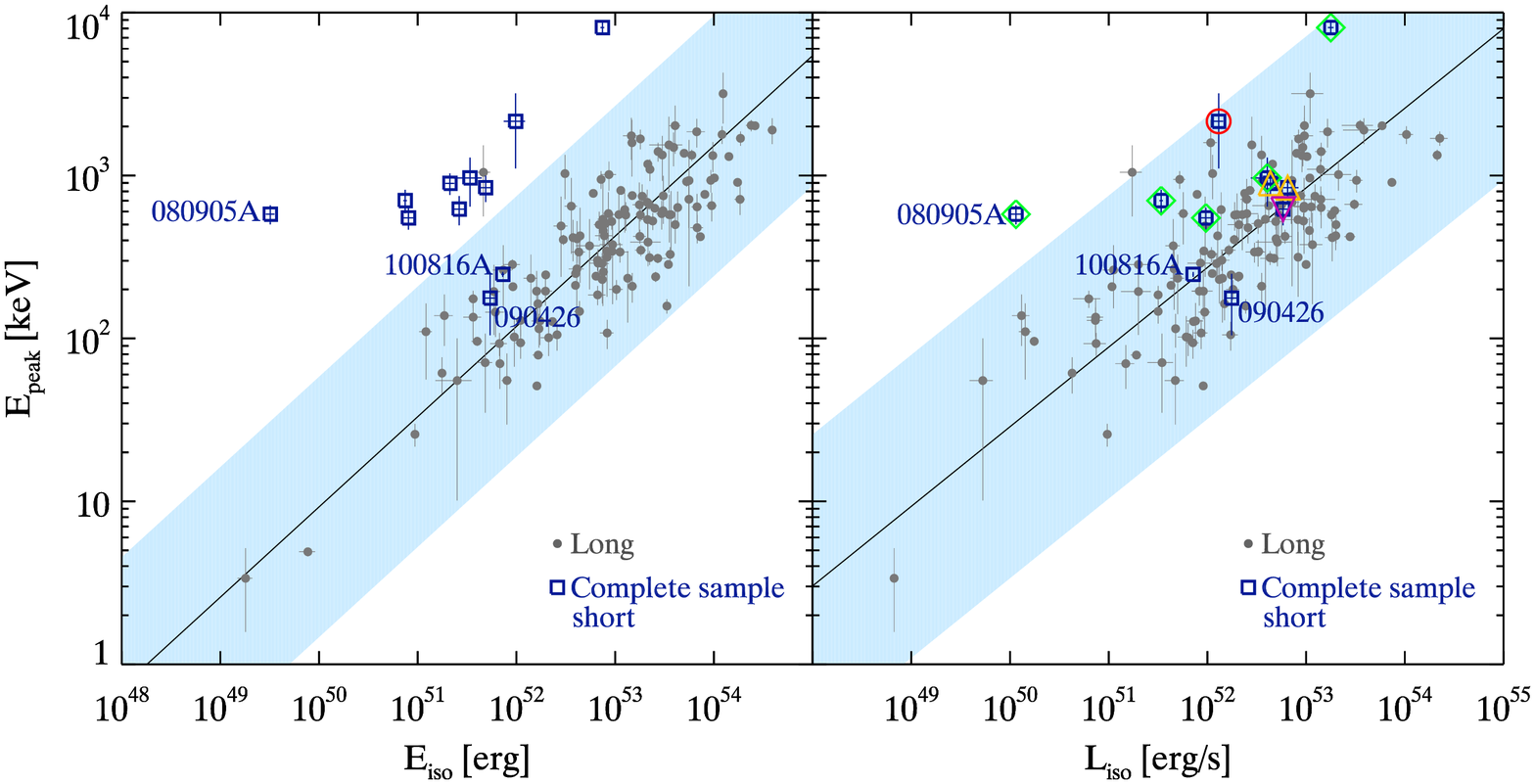}
   \includegraphics[width=162mm]{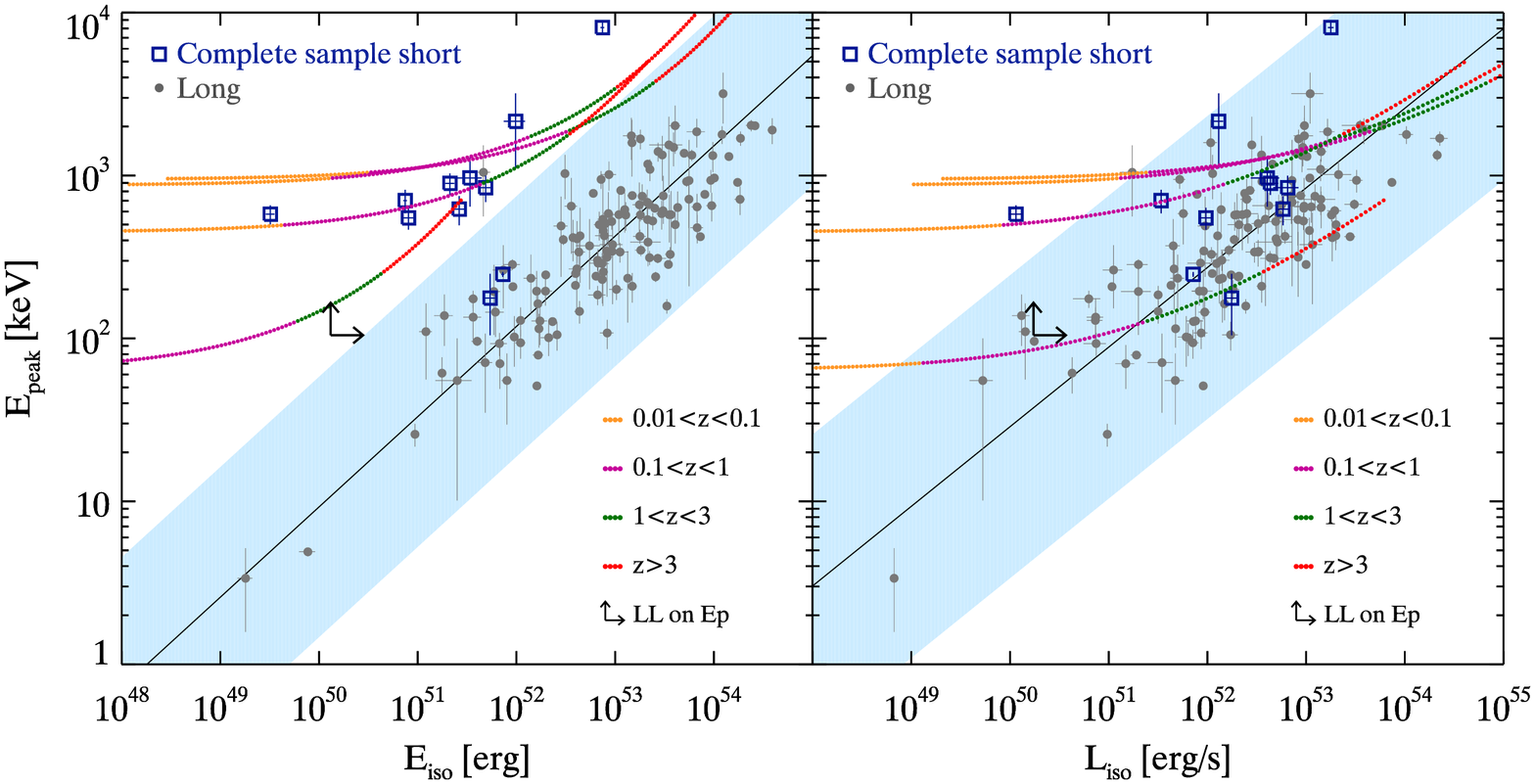}  
\caption{
$E_\mathrm{peak}-E_{\mathrm{iso}}$ ({\it left panels}) and $E_\mathrm{peak}-L_{\mathrm{iso}}$ ({\it right panels}) correlations valid for LGRBs (dots; data taken from Nava et al. 2012). The power-law best fit is shown as a solid dark line. 
The shaded region represents the $3\sigma$ scatter of the distribution. SGRBs of our {\it complete} sample are marked as empty squares. In the lower panels the consistency of the two
correlations of SGRBs with unknown redshift is shown. The test is performed by varying the redshift from 0.01 to 10. Different colours refers to different ranges of redshift (see
legend). Two possible LGRBs belonging to our {\it complete} sample (GRB\,090426 and GRB\,100816A) and a possible outlier of the $E_\mathrm{peak}-L_{\mathrm{iso}}$ correlation (GRB\,080905A) are also marked.
}
       \label{fig:amati_yone}
\end{figure*}

We checked the consistency with the $E_\mathrm{peak}-L_{\mathrm{iso}}$ (Yonetoku et al. 2004), and with the $E_\mathrm{peak}-E_{\mathrm{iso}}$ relations (Amati et al. 2002) for all the bursts of our {\it complete} sample with a measured redshift. The results are shown in the upper panels of Fig.~\ref{fig:amati_yone}.

\begin{figure*}
   \centering
   \includegraphics[height=8.0cm]{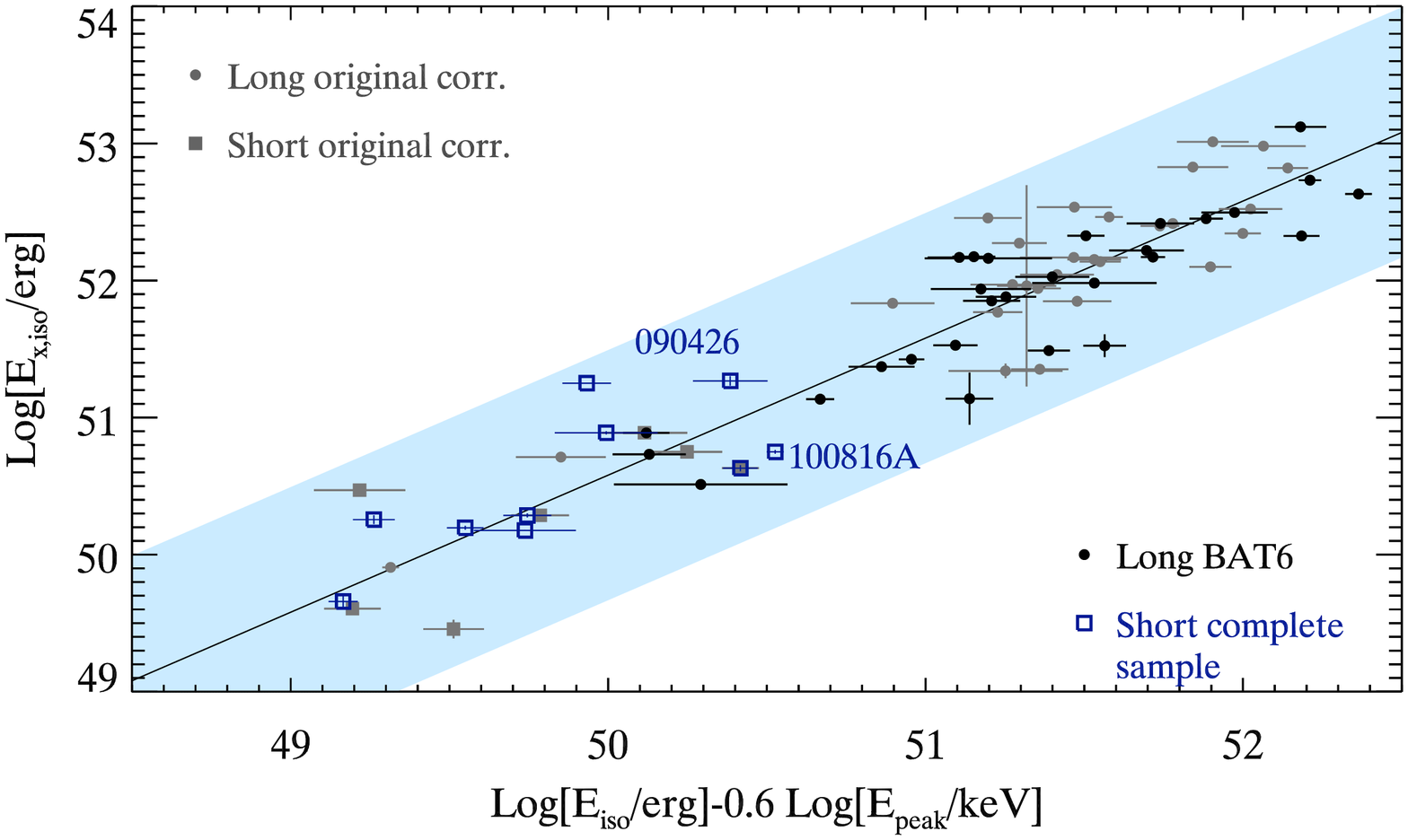}
  \caption{
$E_{iso}-E_\mathrm{peak}-E_{X,iso}$ correlation. The power-law best fit is shown as a solid dark line. The shaded region represents the $3\sigma$ scatter of the distribution. 
SGRBs of our {\it complete} sample are marked as squares. Two possible LGRBs belonging to our {\it complete} sample (GRB\,090426 and GRB\,100816A) are also marked.}
       \label{fig:ExEisoEpk}
\end{figure*}

All events are found to be consistent with the $E_\mathrm{peak}-L_{\mathrm{iso}}$ correlation, which is valid also for LGRBs (Yonetoku et al. 2004; Nava et al. 2012). A fit to the LGRBs with redshift (taken from Nava et al. 2012) together with the SGRBs of the {\it complete} sample with the function $y = 10^Ax^B$ provides a normalization $A = -22.98 \pm 1.81$ and a slope $B=0.49 \pm 0.03$. A significant exception is GRB\,080905A which lies at more than $3\sigma$ from the best fit. This is the event with the lowest redshift of our sample (Table~1) and, consequently, with the lowest values of $L_{\mathrm{iso}}$ and $E_{\mathrm{iso}}$. However, such redshift measurement comes from the association with a spiral galaxy at $z=0.12$, given that the position (known with sub-arcsecond precision) of the optical afterglow of GRB\,080905 falls in the light of this galaxy. As shown in Rowlinson et al. (2010), the proposed host galaxy is seen face-on and the optical afterglow of this burst falls in a region between two spiral arms (and near, but not inside, the bulge). These authors estimate a chance probability alignment of $< 1\%$, a value indeed low but not that implausible. So, either GRB\,080905A is really a peculiar sub-luminous (and sub-energetic) burst, compared to the other events belonging to the SGRB class, or the association with the proposed host galaxy is spurious, leading (likely) to an underestimation of its distance (and, therefore, of its luminosity and energy). 

Even if the SGRBs of our {\it complete} sample are consistent with the $E_\mathrm{peak}-L_{\mathrm{iso}}$ relation, we note that they almost all lie to the left of the best fit of long GRBs.  On the basis of this consideration, Tsutsui et al. (2013) suggested that the $E_\mathrm{peak}-L_{\mathrm{iso}}$ correlation followed by short bursts is 5 times fainter than the same correlation defined by long bursts. The estimate of the peak luminosity depends on the width of the temporal bin $\Delta t$ chosen to rebin the light curve: by choosing smaller and smaller bins, the estimate of $L_{\mathrm{iso}}$ tends to increase. For this reason a more uniform estimate of $L_{\mathrm{iso}}$ should refer to a peak luminosity estimated for all GRBs on the same rest frame temporal bin. In our sample, the peak fluxes have been estimated on different temporal bins, from 4 ms to 1024 ms in the observer frame. In Fig.~\ref{fig:amati_yone} (right panel) we divided our sample in bursts with $\Delta t\ge 1000$ms (red circle), $50\le \Delta t\le 64$ms (green diamonds), $\Delta t=16$ms (yellow triangles) and $\Delta t=4$ms (purple upside down triangle). Bursts with smaller $\Delta t$ systematically tend to have larger $L_{\mathrm{iso}}$, and to be more consistent with the best fit of long bursts. For most long bursts the peak flux has been estimated on a $\Delta t\sim 1$s timescale. Any consideration about the existence of an $E_\mathrm{peak}-L_{\mathrm{iso}}$ correlation for short GRBs and its comparison with the same correlation for long GRBs should take this effect into account. The conclusion can be different depending on the choice of $\Delta t$. When $L_{\mathrm{iso}}$ is estimated on similar timescales both for long and short GRBs, short GRBs lie on the extreme left side of the correlation defined by long bursts. By reducing $\Delta t$ for short bursts (for example for $\Delta t\sim 64$ms) the consistency between long and short in the $E_\mathrm{peak}-L_{\mathrm{iso}}$ plane improves. Our preliminary results might suggest that a consistency between the long and short $E_\mathrm{peak}-L_{\mathrm{iso}}$ correlation can be reached by considering for both classes of events a $\Delta t$ which is a (proper) fraction of their $T_{90}$ (see also Tsutsui et al. 2013).

Concerning the $E_\mathrm{peak}-E_{\mathrm{iso}}$ plane, most of the SGRBs of our sample lie at more than $3\sigma$ from the correlation defined by LGRBs, and systematically on the left with respect to the best fit line of LGRBs (Fig.~\ref{fig:amati_yone}). This suggests for the existence of a SGRB region that has the same slope as the LGRBs relation, but a different normalization (see also Amati 2008; Piranomonte et al. 2008; Ghirlanda et al. 2009; Zhang et al. 2012). Two significant exceptions are GRB\,090426 and GRB\,100816A both placed at the border of the $2\sigma$ confidence region of the relation holding for LGRBs. As discussed in Sect.~2.1, both these events have an uncertain classification. We note, however, that GRB\,090426 is one of the few events of our {\it complete} sample whose prompt emission was observed by the {\it Swift} satellite only. Given the relatively limited energy range of the BAT telescope (15-150 keV) and to the lack of detection by other high-energy satellites, the $E_\mathrm{peak}$ and $E_{\mathrm{iso}}$ for this event were obtained by fitting the BAT spectrum with a Band function with photon indexes fixed to $\alpha=-1$ and $\beta=-2.3$ (see Table~\ref{tab_Eiso}). The same fit performed leaving $\alpha$ and $\beta$ free to vary, only provides a lower limit to $E_\mathrm{peak}$ (with the rest frame $E_\mathrm{peak} > 126$ keV and $E_{\mathrm{iso}} = [2.5-9.2] \times 10^{51}$ erg; Antonelli et al. 2009), thus leaving some uncertainty about the position of this event on the $E_\mathrm{peak}-E_{\mathrm{iso}}$ plane. The situation is different for GRB\,100816A, whose prompt emission was observed by the {\it Fermi} and {\it Konus-Wind} satellites in the $10-2000$ keV energy band, providing a firm estimate of its prompt spectral properties (Fitzpatrick 2010; Golenetskii et al. 2010). In this case, the consistency of this event with the $E_\mathrm{peak}-E_{\mathrm{iso}}$ relation, coupled with its $T_{90}$ of 2.9 s, points toward a classification as a LGRB. Excluding the events whose classification or redshift is uncertain, a fit to the SGRBs of the {\it complete} sample in the $E_\mathrm{peak}-E_{\mathrm{iso}}$ plane with the function $y = 10^Ax^B$ provides a normalization $A = -28.01 \pm 2.91$ and a slope $B=0.60 \pm 0.06$. As a comparison, as reported in Nava et al. (2012), the same fit performed on the total sample of LGRBs provides $A = -26.74\pm 1.13$ and $B=0.55 \pm 0.02$.

Four GRBs (060313, 061201, 090515 and 130515A) have well known spectral properties but unknown $z$. For these GRBs we tested their consistency with the Amati and Yonetoku correlations by varying the redshift from 0.01 to 10 (Fig.~3, bottom panel). Independently of the chosen redshift, they are inconsistent with the Amati correlation (only GRB 061201 can by marginally consistent if its redshift is larger than $\sim2$). For a fiducial redshift lower limit $z>0.1$ they are all consistent with the Yonetoku relation. For GRB\,080123 the redshift has been measured, but only a lower limit on $E_\mathrm{peak}$ and, consequently on $E_{\mathrm{iso}}$ can be estimated from the spectral analysis (arrows in Fig.~3, bottom panel).
For GRB\,080503 it was not possible to test its consistency with the correlations since the spectrum is well described by a power law function and the redshift is unknown.

\begin{figure}
   \centering
   \includegraphics[angle=0,width=84mm]{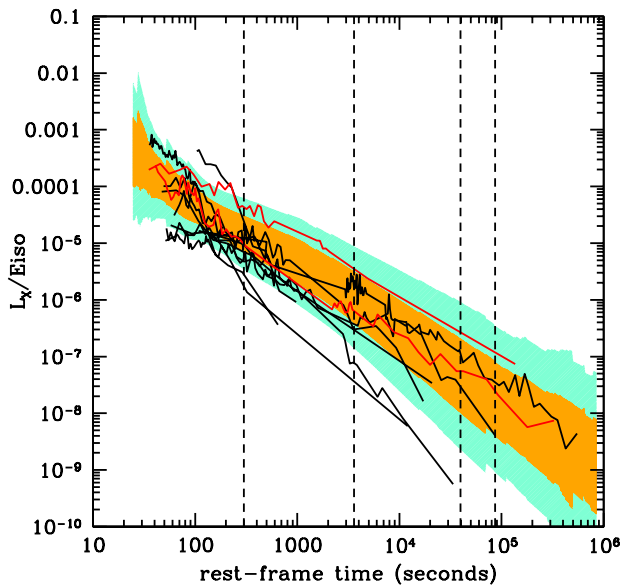}
  \caption{Best fit of the X-ray luminosity light curves of the SGRBs with redshift of our {\it complete} sample normalized to their $E_{\mathrm{iso}}$. The X-ray luminosities were computed for each GRB in the common rest frame $2-10$ keV energy band following the precedure described in Sec. 3.2.2. The rest frame times at which we computed $L_X-E_{\mathrm{iso}}$, $L_X-E_\mathrm{peak}$ and $L_X-L_{\mathrm{iso}}$ correlations are marked with vertical dashed lines. The dark (light) shaded area represent the $1\sigma$ ($2\sigma$) scatter of the same plot obtained for the LGRBs of the BAT6 sample (D'Avanzo et al. 2012). 
}
       \label{fig:Eiso_Lx}
\end{figure}

All the events of our {\it complete} sample are consistent with the SGRB region of the three parameter $E_{iso}-E_\mathrm{peak}-E_{X,iso}$ correlation (Bernardini et al. 2012 and Margutti et al. 2013), with the two debated SGRBs 090426 and 100816A lying close to the region defined by LGRBs (Fig.~\ref{fig:ExEisoEpk}).

\subsubsection{Prompt-afterglow correlations}

In Fig.~\ref{fig:Eiso_Lx} we show the X-ray light curves of the SGRBs of the {\it complete} sample with redshift normalized to their $E_{\mathrm{iso}}$. The distribution of the $E_{\mathrm{iso}}$-normalized X-ray light curves for the LGRBs of the BAT6 sample (taken from D'Avanzo et al. 2012) is also represented in the background for comparison. This plot shows that the $E_{\mathrm{iso}}-$normalized X-ray light curves of long and short GRBs are rather clustered, with an intrinsic scatter that changes with the rest frame time.  With the aim of investigating such evolution in time between the prompt and X-ray afterglow emission, different correlations ($L_X-E_{\mathrm{iso}}$, $L_X-L_{\mathrm{iso}}$ and $L_X-E_\mathrm{peak}$) were tested for the SGRBs of the {\it complete} sample at four different rest frame times. Following the procedure described in D'Avanzo et al. (2012), the early X-ray afterglow luminosity was measured at $t_{rf}= 5$ min and at $t_{rf}= 1$ hr, while the late time afterglow flux was measured at $t_{rf}= 11$ hr and $t_{rf}= 24$ hr (Fig.~\ref{fig:Eiso_Lx}).

\begin{table*}
   \centering
\caption{Correlation fits and coefficients for the prompt vs. early afterglow emission rest frame properties of the {\it complete} sample. Data distributions were fitted with the function $y = 10^Ax^B$ (see Sect. 2.1 for details).}
\begin{tabular}{cccccc}
\hline
Correlation	 & A		     & B	       & $r$  & $P_{null}$	       & Dispersion \\
                 & 		     & 		       &      & 		       &	    \\ \hline
$L_{X,5}$ $vs.$ $E_{\mathrm{iso}}$    &  $-5.24 \pm 6.67$ & $1.00 \pm 0.09$ & 0.67 & $7.19 \times 10^{-2}$  & 0.300   \\
$L_{X,1}$ $vs.$ $E_{\mathrm{iso}}$    &  $-19.24 \pm 18.48$ & $1.25 \pm 0.25$ & 0.43 & $2.97 \times 10^{-1}$  & 0.577   \\
$L_{X,5}$ $vs.$ $L_{\mathrm{iso}}$    &  $ 0.00 \pm 16.07$ & $0.89 \pm 0.20$ & 0.40 & $3.50 \times 10^{-1}$  & 0.622   \\
$L_{X,1}$ $vs.$ $L_{\mathrm{iso}}$    &  $-23.14 \pm 27.06$ & $1.30 \pm 0.33$ & 0.15 & $7.36 \times 10^{-1}$  & 0.775   \\
$L_{X,5}$ $vs.$ $E_\mathrm{peak}$   &  $41.62 \pm 19.83$ & $1.59 \pm 0.55$ & 0.51 & $2.13 \times 10^{-1}$  & 0.463   \\
$L_{X,1}$ $vs.$ $E_\mathrm{peak}$   &  $38.95 \pm 27.71$ & $1.98 \pm 1.06$ & 0.15 & $7.44 \times 10^{-1}$  & 0.631   \\ \hline
\hline
\end{tabular}
\label{tab:log_corr_rf}
\end{table*}

\begin{figure*}
    \centering
    \includegraphics[height=8.0cm,angle=0]{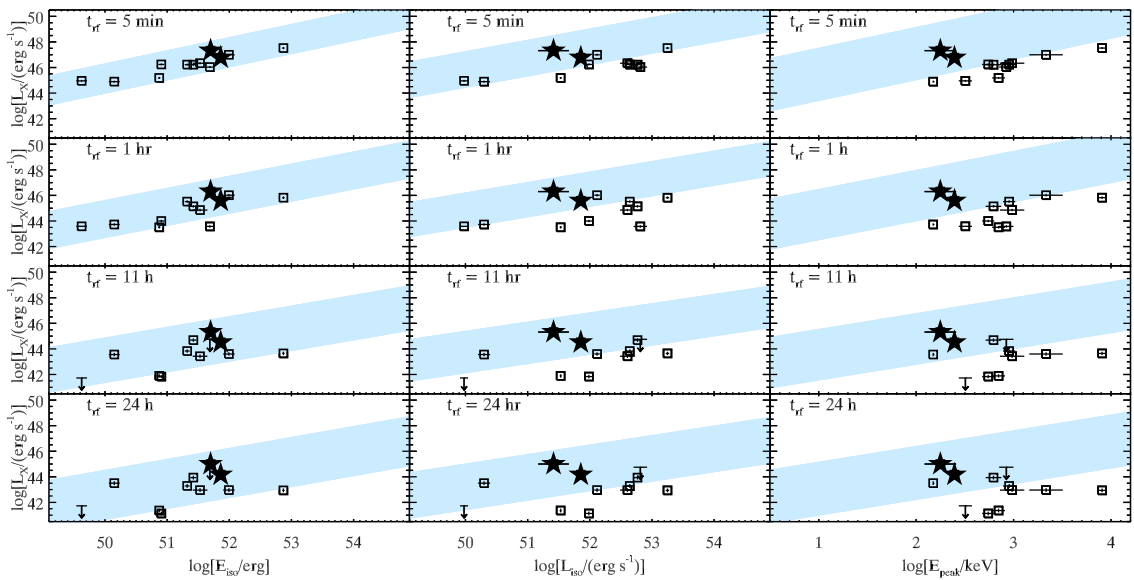}
    \caption{The $L_X-E_{\mathrm{iso}}$, $L_X-L_{\mathrm{iso}}$ and $L_X-E_\mathrm{peak}$ correlations studied in the present paper for the {\it complete} sample of SGRBs at different rest-frame times (stars). GRB\,090426 and GRB\,100816A, found to
    be marginally consistent with the $E_\mathrm{peak}-E_{\mathrm{iso}}$ correlation, are marked with stars. As a comparison, the $3\sigma$ scatter for the correlations found for the LGRBs of the BAT6 sample is also shown (shaded area; from D'Avanzo et al. 2012).}
       \label{fig:correlations}
\end{figure*}

As a general trend, we note that the afterglow X-ray luminosity at early times ($t_{rf}=5$ min) correlates with the prompt emission quantities $E_{\mathrm{iso}}$, $L_{\mathrm{iso}}$ and $E_\mathrm{peak}$ with null probabilities of $10^{-2}-10^{-1}$ and dispersion $\sim 0.3-0.6$ dex. At later times 
($t_{rf}=1$, 11 and 24 hr) the scatter increases and these correlations become much less significant. The early time prompt-afterglow correlation we find (Table~\ref{tab:log_corr_rf}) are rather similar to the same correlations found for the BAT6 sample of LGRBS (D'Avanzo et al. 2012). In particular, the early time $L_X-E_{\mathrm{iso}}$ correlations for short and long GRBs from the two samples have the same slope, with the SGRBs lying on the faint end of the correlation (in agreement with what found by Nysewander et al. 2009). To compare them qualitatively, we show in Fig.~\ref{fig:correlations} the time resolved prompt-afterglow correlations for the SGRBs of our {\it complete} sample and for the LGRBs of the BAT6 sample. Concerning the $L_X-L_{\mathrm{iso}}$  plane we note that at all times, assuming the same $L_{\mathrm{iso}}$, SGRBs have on average lower X-ray luminosity with respect to LGRBs. However, we note that the systematics in the procedure of $L_{\mathrm{iso}}$ estimate for SGRBs discussed in Sect. 4.2.1 for the $E_\mathrm{peak}-L_{\mathrm{iso}}$ relation might amplify this effect. Similarly, compared to LGRBs, for a given $E_\mathrm{peak}$, SGRBs are found at an average lower X-ray luminosity. Considering the correlation found between $L_X$ and $E_{\mathrm{iso}}$, such a result can be considered a natural consequence of the inconsistency of SGRBs with the $E_\mathrm{peak}-E_{\mathrm{iso}}$ correlation (Fig. 3, upper panel). Another feature arising from the comparison of the short and long GRB samples is that the SGRBs prompt-afterglow correlations show a scatter much larger than what observed for LGRBs, especially at late times ($t-t_0 \geq 1$ hr). Although this effect can be related to the lower number of data points in the SGRB sample, it may suggest that the SGRBs X-ray luminosity is not a proxy for $E_{\mathrm{iso}}$ and $L_{\mathrm{iso}}$ as good as it is found to be for the LGRBs (D'Avanzo et al. 2012). Such phenomenon might be indicative of a lower radiative efficiency of the central engine related to the different progenitor or be the consequence of a different circumburst medium. 

Finally, we note from Fig.~\ref{fig:correlations} that the two debated short GRB\,090426 and GRB\,100816A always fall in the LGRBs region of the different correlation planes. While this can be considered as a further evidence against the classification of GRB\,100816A as a SGRB, we stress that the same reasoning cannot be applied safely to GRB\,090426 given the poor constraints on its prompt emission spectrum (see Sect. 4.2.1).

\subsubsection{Redshift distribution}

The redshift distribution of SGRBs can provide an indirect tool to constrain the nature of their progenitors and discriminate among the evolutionary channels. The redshift distribution of merger events of dynamically formed double compact object systems is expected to be different from that of primordial binaries. In particular, given the relatively short delay between formation and merging ($< 1$~Gyr), SGRBs originated by the ``fast'' primordial channel (Sect. 4.1) should have a redshift distribution which broadly follow that of the star formation, especially at low redshift. 

The average (median) redshift for the SGRBs of our {\it complete} sample is $<z>=0.85$ (0.72). This value is  higher than the one obtained by Fong et al. (2013) by considering the whole {\it Swift} SGRB sample\footnote{Including GRB\,050709, discovered by the {\it HETE-II} satellite} ($<z> \sim 0.5$) while it is in agreement with the mean redshift ($<z>=0.72$) reported by Rowlinson et al (2013) for their SGRB sample limited to the events with $T_{90} \leq 2$s (which is thus excluding all SGRB with extended emission). Indeed, the average redshift of our {\it complete} sample is consistent with the expected peak for the redshift distribution of SGRBs originated by the primordial formation channel (Salvaterra et al. 2008). 

%%%%%%%%%%%%%%%%%%%%%%%%%%%%%%%%%%%%%%%%%%%%%%%%%%

In the primordial binary scenario, the intrinsic formation rate, defined as the
number of bursts per unit time and unit comoving volume at redshift
$z$, $\Psi_{\rm SGRB}(z),$ is given by the convolution of the massive
star binary formation rate and a delay time distribution function
$f_{\rm F}(t)$. The delay is the time interval between the formation
of the massive star binary and the merging of the compact objects.  
The formation rate is assumed to be
proportional to the cosmic star formation rate, $\dot{\rho}_\star$,
parametrized as in Hopkins \& Beacom (2006). 
$\Psi_{\rm SGRB}(z)$ is then given by:

\begin{equation}
\Psi_{\rm SGRB}(z) \propto \int_z^{\infty}dz' \frac{dt}{dz'}(z') 
\dot{\rho}_\star(z') f_{\rm F}[t(z)-t(z')],
\end{equation}
where $t(z)$ is the age of the Universe at redshift $z$. 
For the delay time distribution function $f_{\rm F}(t)$, we adopt the
 simple expression: $f_{\rm F}(t) \propto t^{n}$ with $n=-1$, as suggested 
from an updated analysis of double compact object mergers
performed using population synthesis methods (Portegies Zwart \& Yungeson 
1998; Schneider et al. 2001; Belczynski et al. 2006; O'Shaughnessy et al. 2008). 
However, some uncertainty remains on the exact 
value of the $n$ exponent (Berger 2014, and references therein). For a comparison with the data, we compute the observed distribution of SGRBs for three different values of $n$, namely $n=-1.5$, $n=-1$ and $n=-0.5$ (see also Behroozi, Ramirez-Ruiz \& Fryer 2014), with characteristic delay times varying from $\sim 20$ Myr to $\sim 10$ Gyr. Given that a $t^{-1}$ distribution would diverge as $t$ get close to zero, we will consider an initial time cutoff of 10 Myr. As noted by Behroozi, Ramirez-Ruiz \& Fryer (2014), the choice of the time cutoff has little effect on the mean time delay value. 

\begin{figure}
   \centering
   \includegraphics[angle=0,width=84mm]{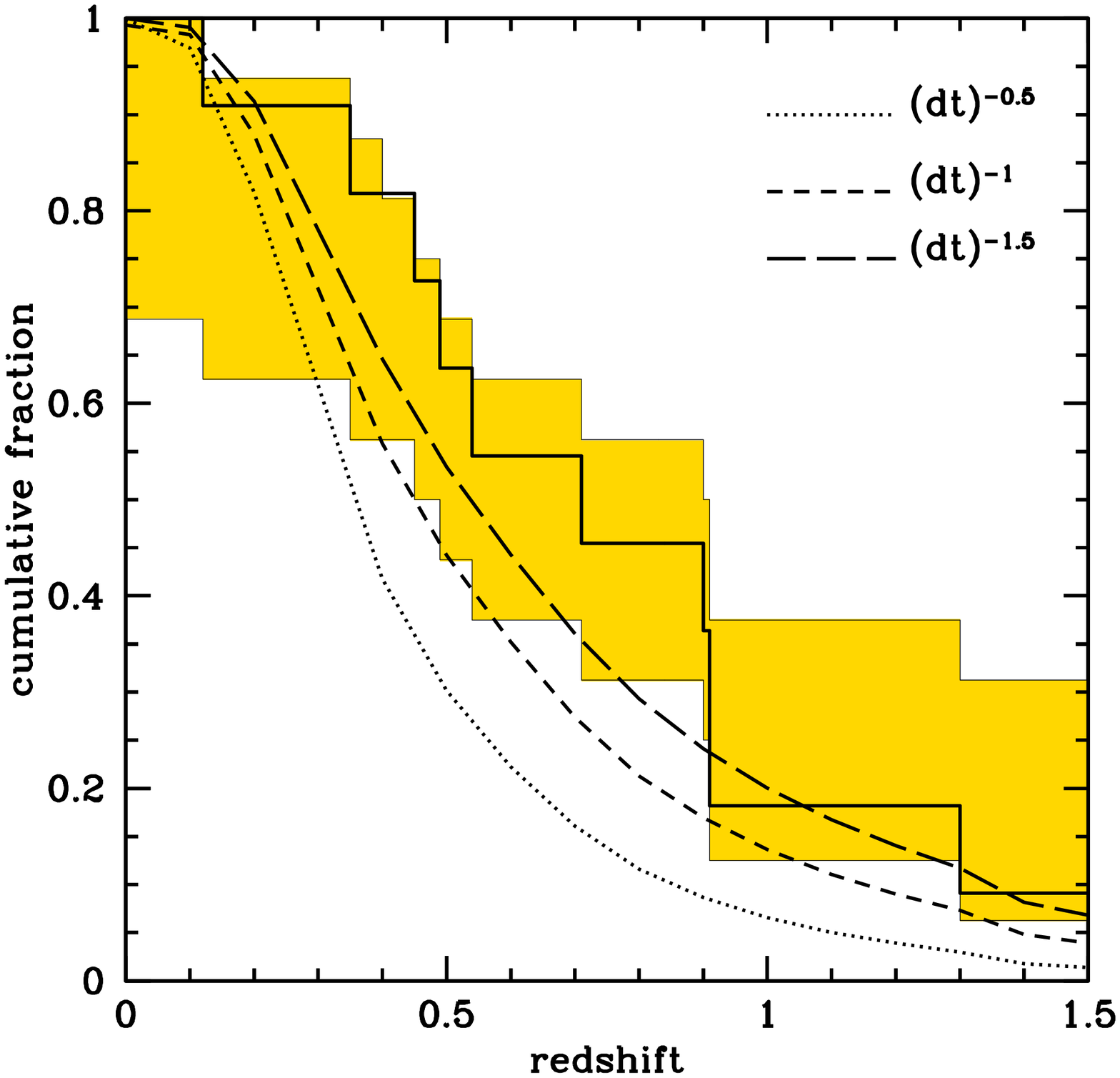}
  \caption{Redshift distribution of our {\it complete} sample of
    SGRBs. The shaded are takes into account the uncertainties due to
    the lack of redshift measurement for five bursts in the
    sample. Model results for $n=-1.5$, -1, and -0.5 are shown with
    the long-dashed, short-dashed and dotted line, respectively. In
    computing the expected redshift distribution for the different
    model we apply the same photon flux cut, $P_{64}\ge 3.5$ ph s$^{-1}$
    cm$^{-2}$ in the
    {\it Swift-}BAT 15--150 keV band, used in the definition of our {\it complete} sample.
}
       \label{fig:z_distrib}
\end{figure}

The observed photon flux, $P$, in the energy band $E_{\rm min}<E<E_{\rm max}$, emitted by an isotropically radiating source at redshift $z$ is:

\begin{equation}
P=\frac{(1+z)\int^{(1+z)E_{\rm max}}_{(1+z)E_{\rm min}} S(E) dE}{4\pi d_L^2(z)},
\end{equation}

\noindent
where $S(E)$ is the differential rest--frame photon luminosity of the
source, and $d_L(z)$ is the luminosity distance.  We define the isotropic equivalent intrinsic 
burst luminosity as $L=\int^{10000\,\rm{keV}}_{1\,\rm{keV}} S(E)\, E\, dE$ and, to describe the
typical burst spectrum, we adopt a Band function with photon indexes
fixed to $\alpha=-0.6$ and $\beta=-2.3$ and rest-frame peak energy
$E_{\rm peak} = 337\;{\rm keV} (L_{\rm iso}/2\times 10^{52}\;\rm{erg \, s}^{-1})^{0.49}$, assuming that the $E_{\rm peak}-L_{\rm iso}$ is valid for SGRBs (see Sect. 4.2.2). 

Given a normalized SGRB luminosity function, $\phi(L)$, the observed rate of bursts with peak flux between $P_1$ and $P_2$ is:

\begin{eqnarray}
\frac{dN}{dt}(P_1<P<P_2)&=&\int_0^{\infty} dz \frac{dV(z)}{dz}
\frac{\Delta \Omega_s}{4\pi} \frac{k_{\rm SGRB} \Psi_{\rm SGRB}(z)}{1+z}
\nonumber \\ & \times & \int^{L(P_2,z)}_{L(P_1,z)} dL^\prime
\phi(L^\prime),
\end{eqnarray}

\noindent
where $dV(z)/dz=4\pi c\, d_L^2(z)/[H(z)(1+z)^2]$ is the comoving volume
element, and $H(z)=H_0 [\Omega_M
(1+z)^3+\Omega_\Lambda+(1-\Omega_M-\Omega_\Lambda)(1+z)^2]^{1/2}$.
$\Delta \Omega_s$ is the solid angle covered on the sky by the survey,
and the factor $(1+z)^{-1}$ accounts for cosmological time dilation.
$\Psi_{\rm SGRB}(z)$ is the comoving burst formation rate normalized to
unity at $z=0$ as computed in Eq.~3 and $k_{\rm SGRB}$ is the SGRB
formation rate at $z=0$. Finally, we assume that the luminosity function of short GRBs is well described
by a single power-law of power-index $\xi$ for luminosity in the range
$10^{49}-10^{55}$ erg s$^{-1}$. Following Salvaterra et al. (2008), we obtain the value of $\xi$ by
fitting the differential photon flux distribution of SGRBs  with $T_{90}<2$ s 
detected by BATSE in the 50-300 keV band (Paciesas et al. 1999).
%restrict our fitting procedure to $P_{64}\ge 1$ ph s$^{-1}$ cm$^{-2}$
%or which the detector response is efficient, and adopt a effective field of
%view for BATSE, defined as $(\Delta \Omega_s/4\pi) t_{\rm obs},$  
%of 1.8 yr. The best fit parameters are given in Table~1. 
We obtain a reasonable fit to BATSE data in all formation scenarios
with $\xi=-2.17\pm 0.08$, $-2.19\pm 0.08$, and $-2.22 \pm 0.08$ (errors at 1$\sigma$ confidence level) for $n=-1.5$, $-1$ and
$-0.5$, respectively.

We then compute the expected redshift distribution of short GRBs
matching the selection criteria of our {\it complete} sample,
i.e. those  with photon fluxes $P_{64}\ge 3.5$ ph s$^{-1}$ cm$^{-2}$ in the
15--150 keV band of {\it Swift-}BAT (see Sect.~2). We note that having a 
flux-limited sample, with a clear photon flux cut, is an asset for the comparison 
with the model predictions. 
Model results computed for the primordial formation channel with
different time delay distributions are shown in
Fig.~\ref{fig:z_distrib}. The shaded area accounts for the observed
cumulative redshift distribution including the uncertainties due the
lack of a secure redshift determination for $\sim 30\%$ of the SGRBs
of the {\it complete} sample. From the plot it is clear that the model
with $n=-0.5$ can be firmly discarted. A model with $n=-1$ is
still acceptable (although at the lower limit of the observed redshift
distribution), while the model with a time delay distribution $f_{\rm F}(t)
\propto t^{-1.5}$ looks to be favoured in accounting for the observed
redshift distribution of the SGRBs of our sample, suggesting that they 
are mainly originated by primordial double compact object
systems merging in a relatively short time. This differs from the
results of a similar analysis carried on for the BAT6 sample of long
GRBs: to account for their observed
redshift distribution, the rate of long
GRB formation has to increase with redshift on the top of the known
evolution of the SFR density as $(1+z)^{1.7\pm 0.5}$ (Salvaterra et
al. 2012).

%%%%%%%%%%%%%%%%%%%%%%%%%%%%%%%%%%%%%%%%%%%%%%%%%%

Finally, we note that a significant contribution of SGRBs with dynamical origin would require a lower mean redshift (Salvaterra et al. 2008; Guetta et al. 2008), suggesting that the contribution of this formation channel to the SGRBs should be negligible and/or limited to the faintest events (which are not included in our flux-limited sample). On the other hand one can conclude that dynamically formed SGRBs should be intrinsically sub-energetic given that, occurring at an average lower redshift, they should show a relatively high observed flux (and, in principle, fall in a flux-limited sample like the one presented in this paper). A tentative estimate of the fraction of SGRBs with dynamical origin in our sample is given in the next subsection.

\subsubsection{Environment }

\begin{figure}
    \centering
    \includegraphics[width=84mm,angle=0]{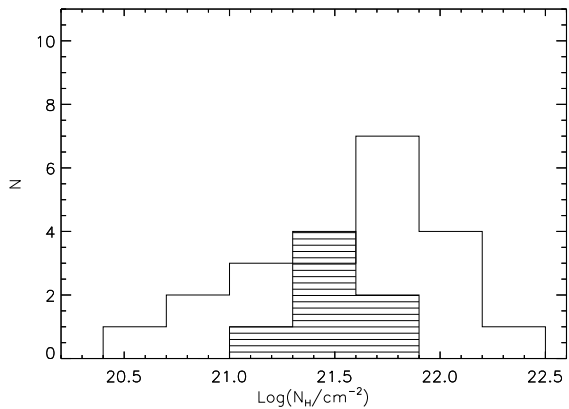}
    \caption{Distribution of the intrinsic X-ray absorbing column densities for the SGRBs of the {\it complete} sample (filled histogram) and of the LGRB with $z<1.3$ of the BAT6
    sample (data taken from Campana et al. 2012).}
       \label{fig:nh}
\end{figure}

The distribution of the intrinsic X-ray absorbing column densities of the SGRB of our {\it complete} sample has a mean value $\mathrm{log\,N_H} (z) = 21.5$ and a standard deviation $\sigma _{\mathrm{log\,N_H}(z)} = 0.2$ (Fig.~\ref{fig:nh}). We compared our values with the distribution of the intrinsic X-ray absorbing column densities of the {\it complete} sample of LGRBs presented in Campana et al. (2012). In order to make a proper comparison, we considered only those LGRBs whose redshift is lower or equal to $z=1.3$ (which is the highest redshift recorded in our SGRB {\it complete} sample) obtaining a mean $\mathrm{log\,N_H} (z) = 21.6$ and a standard deviation $\sigma _{\mathrm{log\,N_H}(z)} = 0.4$ (Fig.~\ref{fig:nh}). When compared in the same redshift bin, the distribution of the intrinsic X-ray absorbing column densities of the two sample are fully consistent. A two-sample Kolmogorov-Smirnov (K-S) test gives a probability of 34\%, likely indicating that the two distributions are drawn from the same population. Such result is in agreement with the findings (obtained following a similar procedure) by Kopac et al. (2012) and Margutti et al. (2013).

Although this result can be intepreted as the evidence of a common environment for long and short GRBs, we caution that the intrinsc X-ray $N_H$ might be a good proxy of the GRB host galaxy global properties but not for the specific properties of the circumburst medium. Furthermore, the possibility that gas along the line of sight in the diffuse intergalactic medium or intervening absorbing systems can contribute to the absorption observed in the X-ray emission of GRBs has to be taken into account (Behar et al. 2011; Campana et al. 2012; Starling et al. 2013). However, such effect is expected to dominate at $z \geq 3$, while at lower redshifts, comparable to the values found for our {\it complete} sample, the absorption within the GRB host galaxy is expected to dominate (Starling et al. 2013). For LGRBs, the massive star progenitor is expected to significantly enrich the surrounding environment with metals (whose X-ray $N_H$ is a proxy) before the collapse with its stellar wind. Alternatively, it has been recently proposed that the Helium in the H II regions where the burst may occur is responsible for the observed X-ray absorption in LGRBs (Watson et al. 2013). Under these hypothesis, a high intrinsic X-ray $N_H$, can be interpreted as the evidence of a dense circumburst medium.  Something similar can happen for SGRBs, under the condition that a short time (of the order of Myrs) separates the supernova explosions which gave origin to the compact objects in the primordial binary system progenitor and its coalescence, with the result that the burst would occur inside its host galaxy and near its star forming birthplace (Perna \& Belczynski 2002). Such formation channel of ``fast merging'' primordial binaries is in agreement with the observed redshift distribution of our {\it complete} sample discussed above.  Indeed, the only case for which combined X-ray and optical afterglow spectroscopy could be performed for a genuine SGRBs (GRB\,130603B, which is included in our sample), provided evidence for a progenitor with short delay time or a low natal kick (de Ugarte Postigo et al. 2013).

SGRBs originated by double compact object systems which experienced a large natal kick or which are dynamically formed in globular clusters are expected to be associated with a low-density environments. As shown in Table~4, four SGRBs of the {\it complete} sample have only upper limits on the intrinsic  X-ray $N_H$. Among these, GRB\,100625A is the only event whose upper limit is significantly below the average $N_H$ of the distribution. Assuming that such limit is indicative of a low-density circumburst medium, we can estimate that at least 10\% of the events of our sample are originated by coalescing binaries formed via the dynamical channel (or having experienced a large natal kick). Furthermore, three (out of five) events of the {\it complete} sample missing a robust redshift measure (GRB\,061201, GRB\,080503, GRB\,090515) have no detected host galaxy coincident with the afterglow position in spite of the precise (sub-arcsecond) location and the deep magnitude limits (Berger 2010 and references therein). As discussed in Berger (2010) ``hostless'' SGRBs may lie at moderately high redshifts $z > 1$, and have faint hosts, or represent a population where the progenitor has been kicked out from its host or is sited in an outlying globular cluster. A statistical study carried out recently by Tunnicliffe et al. (2014) pointed out that the proximity of these events to nearby galaxies is higher than is seen for random positions on the sky, in contrast with the high redshift scenario. Following this interpretation, up to 4 events (25\% of the SGRBs of the {\it complete} sample)\footnote{Namely, GRB\,100625A and the three ``hostless'' bursts.} might have occurred in low-density environments. 

Finally, we remark that the {\it complete} sample presented in this paper is built by selecting the events with a bright prompt emission that, according to the standard GRB model, is independent on the type of circumburst environment.

\section{Conclusions and future works}

The statistical study of the rest-frame properties of SGRBs gives the best opportunity to characterize the physics of these events, although such studies are often biased by the fact that almost 3/4 of GRBs are lacking a secure redshift measurement. In this paper, we overcome this problem working with a carefully selected sample of {\it Swift} SGRBs having a completeness in redshift of $\sim 42\%$ which increases up to $\sim 70\%$ by considering the events with the brightest $\gamma-$ray prompt emission. From the study of the observer and rest-frame properties of this sample we obtained the following main results:

\noindent - The SGRBs X-ray afterglow fluence correlates linearly with the prompt emission fluence at all times, with no difference for the SGRBs with extended emission or with a short/long-lived X-ray afterglow. 

\noindent - The percentage of {\it Swift} SGRBs lacking an X-ray afterglow detection at early times is more than 6 times higher than for LGRBs. These events lay in the faint end of the $\gamma-$ray fluences distribution suggesting that their lack of an X-ray afterglow is likely due to the intrinsic faintness of the prompt emission, although a high redshift origin for these events cannot be excluded.

\noindent - Compared to bright LGRBs (BAT6 sample) the spectral hardness of the SGRBs of the {\it complete} sample seems to be mainly due to a harder low energy spectral component present in short bursts, rather than to a higher peak energy. 

\noindent - All the SGRBs of our {\it complete} sample are consistent with the $E_\mathrm{peak}-L_{\mathrm{iso}}$ and the $E_{iso}-E_\mathrm{peak}-E_{X,iso}$ correlations, with the significant exception of GRB\,080905A. We note that such event has the lowest values for $E_{\mathrm{iso}}$ and $L_{\mathrm{iso}}$ among the burst of our sample. We speculate that the redshift of this GRB might be higher than the value inferred from its association with a nearby ($z=0.122$) spiral galaxy observed edge-on. Apart from this exception, we found evidence for a $E_\mathrm{peak}-L_{\mathrm{iso}}$ correlation followed by SGRBs being sistematically fainter than the correlation defined by LGRBs. Although such finding is intriguing, we caution that it can be affected by the choice of the temporal bin in the estimate of the isotropic peak luminosity for both long and short GRBs. 

\noindent - On the $E_\mathrm{peak}-E_{\mathrm{iso}}$ plane, SGRBs define a region with the same slope measured for the correlation holding for LGRBs but with a different normalization. Two exceptions are GRB\,090426 and GRB\,100816A, both consistent within $2\sigma$ with the $E_\mathrm{peak}-E_{\mathrm{iso}}$ LGRB correlation. 

\noindent - The rest-frame afterglow X-ray luminosity at early times ($t_{rf} = 5$ min) correlates well with the prompt emission $E_{\mathrm{iso}}$ and $L_{\mathrm{iso}}$. At later times ($t_{rf} \geq 1$ hr) these correlations become much less significant. This is at variance with respect to what observed for LGRBs, where, although with increasing scatter from early to late times, the $L_{X}-E_{\mathrm{iso}}$ and $L_{X}-L_{\mathrm{iso}}$ correlations hold at all times. Such effect might be indicative of a lower radiative efficiency of the SGRB central engine related to the different progenitor or be the consequence of a different circumburst medium for long and short GRBs.

\noindent - In light of its duration, prompt emission and X-ray afterglow properties, GRB\,100816A (initially classified as a SGRB) is likely a long-duration event. No firm conclusion can be derived for the classification of GRB\,090426, mainly due to the lack of strong constraints on the properties of its prompt emission spectrum.

\noindent - The redshift distribution of the SGRBs of our sample has a mean value of $z=0.85$. No evidence for a different environment for long and short GRBs can be derived from the rest-frame X-ray column densities. When compared on the same redhift bin, the distributions for long and short GRBs are fully comparable. Both these results are consistent with the scenario of ``primordial binary'' progenitors, with short coalescence times. However, a minor contribution (10\%--25\%) of dynamically formed (or with large natal kicks) compact binaries progenitors cannot be excluded.

\smallskip

The {\it complete} sample of SGRBs presented in this paper consists of 16 events observed by the {\it Swift} satellite over 8.5 years. As can be evinced from Table~1, the efficiency of redshift measurements almost doubled (from $\sim 30\%$ to $\sim 60\%$) in the last 5.5 years (since 2008). This is likely due to the improving procedures of automatic follow-up from the ground and to the advent of effective facilities like GROND (Greiner et al. 2008; Nicuesa Guelbenzu et al. 2012). As a consequence of this, within a few years of further {\it Swift} operations we will handle a well-selectd sample of  $\sim 30$ SGRBs with a redshift completeness higher than 70\%. With such numbers, the statistics of this SGRB sample will become comparable to, e.g., the LGRB BAT6 sample, opening the possibility to study, e.g., the SGRBs luminosity function and measuring their intrinsic extinction through their optical and near-infrared spectral energy distributions (Salvaterra et al. 2012; Covino et al. 2013).

\section*{Acknowledgments}
This work made use of data supplied by the UK Swift Science Data Centre at the University of Leicester.
We acknowledge the Italian Space Agency (ASI) for financial support through the ASI-INAF contract I/004/11/1.

\label{lastpage}

\end{document}